\documentclass[11pt,a4paper]{article}
\usepackage[T1]{fontenc}
\usepackage[utf8]{inputenc}
\usepackage[english]{babel}
\usepackage{authblk}
\usepackage{amsthm}
\usepackage{color}
\usepackage[colorlinks=true,urlcolor=blue,linkcolor=blue]{hyperref}
\usepackage{graphicx}
\usepackage[bottom]{footmisc}
\usepackage{fullpage}
\usepackage{calc}
\usepackage{amsmath,amssymb}
\usepackage{amsfonts}
\usepackage{mathrsfs}
\usepackage[nottoc]{tocbibind}
\usepackage{graphicx}
\graphicspath{ {images/} }
\usepackage{fancyhdr}
\usepackage{tabularx}
\usepackage{array}
\usepackage{lscape}
\usepackage{array,multirow,makecell}
\setcellgapes{1pt}
\makegapedcells
\usepackage[table]{xcolor}
\usepackage{float}
\usepackage{caption}
\usepackage{subcaption}
\usepackage{pifont}
\usepackage[ruled,vlined,linesnumbered,noresetcount]{algorithm2e}
\usepackage{booktabs}
\usepackage{titlesec}
\usepackage[section]{placeins}
\usepackage{bbm}
\hypersetup{citecolor=blue}
\newcolumntype{R}[1]{>{\raggedleft\arraybackslash }b{#1}}
\newcolumntype{L}[1]{>{\raggedright\arraybackslash }b{#1}}
\newcolumntype{C}[1]{>{\centering\arraybackslash }b{#1}}

\newtheorem{prop}{Proposition}
\newtheorem{defi}{Definition}
\newtheorem{coro}{Corollary}
\newtheorem{rem}{Remark}

\makeatletter
\renewcommand\paragraph{\@startsection{paragraph}{4}{\z@}
  {-3.25ex \@plus -1ex \@minus -0.2ex}
  {2.25ex \@plus .25ex}
  {\normalfont\normalsize\bfseries}}
\renewcommand\subparagraph{\@startsection{subparagraph}{5}{\z@}
  {-3.25ex \@plus -1ex \@minus -0.2ex}
  {2.25ex \@plus .25ex}
  {\normalfont\normalsize\bfseries}}
\def\toclevel@paragraph{4}
\def\toclevel@paragraph{5}
\def\l@paragraph{\@dottedtocline{4}{7em}{4em}}
\def\l@subparagraph{\@dottedtocline{5}{7em}{4em}}
\makeatother

\title{Market Impact: A Systematic Study of Limit Orders \\
\small \textit{(published in Market Microstructure and Liquidity Vol. 3, Nos. 3\&4 2017)}}
\author[*,**]{Emilio Said}
\author[*]{Ahmed Bel Hadj Ayed}
\author[*]{Alexandre Husson}
\author[**]{Frederic Abergel}

\affil[*]{Quantitative Research, Global Markets, BNP Paribas, Paris, France}
\affil[**]{Chaire de Finance Quantitative, Laboratoire de Mathématiques et Informatique pour la Complexité et les Systèmes, CentraleSupélec, Université Paris-Saclay, Gif-Sur-Yvette, France}

\begin{document}

\maketitle

\begin{abstract}
This paper is devoted to the important yet little explored subject of the market impact of \emph{limit orders}. Our analysis is based on a proprietary database of metaorders - large orders that are split into smaller pieces before being sent to the market. We first address the case of \textit{aggressive} limit orders and then, that of \textit{passive} limit orders. In both cases, we provide empirical evidence of a power law behaviour for the temporary market impact. The relaxation of the price following the end of the metaorder is also studied, and the long-term impact is shown to stabilize at a level of approximately two-thirds of the maximum impact. Finally, a \emph{fair pricing} condition during the life cycle of the metaorders is empirically validated.
\end{abstract}


\section{Introduction}

It is a commonly acknowledged fact that market prices move during the execution of a trade - they go up for a large buy order and decrease for a large sell order. This is, loosely stated, the phenomenon known as \emph{market impact}.

The market impact of large trading orders that are split into pieces - better known as \textbf{metaorders} - and executed incrementally through a sequence of orders of smaller sizes is crucial in describing and controlling the behaviour of modern financial markets. Being able to quantify this impact is clearly a question of great relevance when studying the price formation process, and it has also become a major practical issue for optimal trading. Indeed, in order to know whether a trade will be profitable, it is essential to monitor transaction costs, which are directly linked to market impact. Measuring and modelling market impact has therefore become a central question of interest, both for researchers and practitioners, in the field of market microstructure.

Given the importance of the subject, there exists but a few research articles pertaining to the empirical estimation of market impact, mostly due to the scarcity of data. In fact, trades and quotes, or even order book databases are not sufficient to perform the analysis: what is required is a clear identification of metaorders. Metaorders have started being recorded as such in a systematic way only recently, and mostly in proprietary databases that are not readily accessible to academic researchers in the field of market microstructure. The analyses presented in \cite{almgren2005direct} \cite{moro2009market} \cite{gatheral2010no} \cite{toth2011anomalous} \cite{bershova2013non} \cite{bacry2015market} \cite{zarinelli2015beyond} \cite{gomes2015market} essentially cover all that is published about the market impact of large orders.

Although difficult to measure in practice, market impact has been studied from a theoretical point of view. In an economic theory perspective, the information held by investors, which governs their decisions, should have some predictive power over future prices. This point was thoroughly investigated by \cite{farmer2013efficiency}, a paper which we find enlightening and use as reference for the theoretical measurements of market impact. In a related study, \cite{bershova2013non} uses their own proprietary database to perform an empirical analysis of a set of large institutional orders, and validates some predictions of the \cite{farmer2013efficiency} model. Such a comparison will also be performed in the present work.

Our paper is a contribution to this strand of research, with a specific focus on the market impact of \emph{limit orders}. As a matter of fact, market orders are generally not used by institutional investors because of the lack of control they imply. On the contrary, limit orders, whether they are \emph{aggressive} - crossing the spread - or \emph{passive}, form the vast majority of orders that are actually sent to the market during the execution of a large trade. As such, they should be the main subject of interest in a study of market impact. To the best of our knowledge, ours is the first academic study of market impact with an emphasis on limit orders. The statistical results presented in this paper are obtained from a proprietary database consisting of appropriately selected limit orders executed on the European equity market between January 2016 and December 2017. The originality of our approach lies in the fact that the study relies on an algorithmic-based identification and reconstruction of metaorders from the database of all orders.

The paper is organized as follows: Section \ref{market impact history} is a short review of the literature on market impact. Section \ref{definitions} introduces our main definitions of metaorders and market impact measures. Section \ref{empirical} presents our empirical results: They confirm that limit orders behave in agreement with some \emph{stylized} facts already established in the literature, and shed a new light on the influence of passive orders in an execution strategy. Section \ref{square root law} considers the so-called \textit{Square Root Law} regarding to our metaorders. Section \ref{fair pricing} deals with the fair pricing condition concerning our metaorders. Section \ref{conclusion} is a discussion of our results and their implications, including some comparisons with the literature. Appendix \ref{Farmer} recalls for reference the framework and main results of the theoretical, agent-based market impact model developed in \cite{farmer2013efficiency}.

\section{A short review of the market impact literature}
\label{market impact history}

The strategic reasons underlying the incremental execution of metaorders were originally analyzed by \cite{kyle1985continuous}, where a model considering an insider trader with monopolistic information about future prices is introduced. It is shown that the optimal strategy for such a trader consists of breaking its metaorder into pieces and execute them incrementally at a uniform rate, gradually incorporating its information into the price. In Kyle's theory the total impact is a linear function of size, and the price increases linearly with time as the auctions take place. The prediction of linearity also appeared in the work of \cite{huberman2004price}. They showed that, in the case of constant liquidity providing, and in order to prevent arbitrage, the permanent impact must be linear, i.e. incremental impact per share remains constant during the metaorder life. It is however the case that real data contradict these predictions: metaorders do not exhibit linear market impact. In fact, most empirical studies consistently highlight a concave impact, in sharp contrast with the theoretical linear shape. The first relevant empirical study of market impact is \cite{almgren2005direct}, which directly measures the market impact of large metaorders in the US equity market and provides empirical evidence of a concave temporary impact. Later studies \cite{moro2009market} \cite{toth2011anomalous} \cite{bacry2015market} also find a concave market impact, in rough agreement with a square root formula. Following these experimental findings, some theoretical efforts have been made to reconcile Kyle's model with a concave dependence on size, essentially by adding the hypothesis that larger metaorders contain less information per share than smaller ones. \cite{farmer2013efficiency} presents a model enriching Kyle's approach with this concave dependence on size. Whereas Kyle considers a single, monopolistic informed trader, \cite{farmer2013efficiency} introduces several competitive traders receiving a common information signal and then choosing independently the size of the order they submit to an algorithmic execution service. This set-up is close in spirit to the real-life organization of the major players in the market, which operate by setting up an internal market with a stakeholder recovering all the orders before executing them on the external market.

To be even more specific, market impact can be studied from two different perspectives. The first one, introduced in the previous paragraph, addresses the effect of a metaorder being executed on the price formation process. This effect is commonly termed the {temporary market impact}. It is clearly an important explanatory variable of the price discovery and is studied as such in several papers \cite{kyle1985continuous} \cite{hautsch2012market} \cite{doyne2004really}. Temporary market impact is obviously the main source of trading costs, and models based on empirical measurements can be used in optimal trading schemes \cite{almgren2005direct} \cite{gatheral2010no} \cite{lehalle2010rigorous} \cite{almgren2001optimal} \cite{gatheral_schied_2013}, or used by an investment firm in order to understand its trading costs \cite{bershova2013non} \cite{brokmann2015slow} \cite{mastromatteo2014agent}. One common conclusion to the studies is that the temporary market impact is mainly characterized by three components. The first, obvious one is the size of the metaorder, suitably rescaled by a quantity reflecting the traded volume of the security under scrutiny. The daily participation or trading rates capture most of the dynamics of this component. Note that some empirical studies such as \cite{brokmann2015slow} prefer to consider as a scaling factor the participation rate during the metaorder life. This approach presents the advantage to include a duration effect not captured by the daily participation rate, and it will be used in the present study. One must also allow for some dependency on the price uncertainty while the metaorder is executed, and the volatility or the bid-ask spread are typical measures of this uncertainty. Last but not least, it seems necessary to capture the information leakage generated by the metaorder, the number of orders executed during the metaorder or its duration being good proxies.

There is a second, more controversial type of market impact, pertaining to the persistence of a shift in the price after the metaorder is fully executed. This effect is commonly called the \emph{permanent market impact}. Research papers dealing with permanent impact can be separated in two categories. The first one considers the permanent impact as the consequence of a mechanical process. The second ones consider the permanent impact as the trace of new information in the price. Among those who share the mechanical vision of the permanent market impact, there are two approaches. The first picture of \cite{farmer2013efficiency} and \cite{bershova2013non} says that the permanent impact is important and roughly equals to 2/3 of the temporary impact. This is derived from a fair-pricing hypothesis. In the second picture \cite{bouchaud2010price}, there is actually no such thing as a permanent impact, the slow decay of the market impact being the result of the long memory of the order flow. These two approaches are incompatible. In the present paper, the studies performed on the price relaxation seem to advocate in favour of a permanent impact at the approximate two-thirds level, in agreement with the existing empirical literature as well as the fair pricing condition of \cite{farmer2013efficiency}.

\section{Definitions, Algorithm and Market Impact measures}
\label{definitions}

\subsection{Basic Definitions}

Some basic concepts, and the algorithmic definition of a metaorder, are introduced here.

\begin{defi}
A \textbf{limit order} is an order that sets the maximum or minimum price at which an agent is willing to buy or sell a given quantity of a particular stock.
\end{defi}

\begin{defi}
An \textbf{aggressive limit order} is one that instantaneously removes liquidity from the order book by triggering a transaction. An aggressive order crosses the Bid–Ask spread. In other words an aggressive buy order will be placed on the ask, and an aggressive sell order will be placed on the bid.
\end{defi}

A limit order that is not aggressive is termed \textbf{passive}. Passive orders sit in the order book until they are executed or cancelled.

Loosely speaking a \textbf{metaorder} is a large trading order that is split into small pieces and executed incrementally. In order to perform rigorous statistical analyses, a more specific and precise definition of a metaorder is required, and given in Definition \ref{defMetaorder I} below:

\begin{defi}
\label{defMetaorder I}
A \textbf{metaorder} is a series of orders sequentially executed during the same day and having those same attributes:
\begin{itemize}
    \item \textcolor{blue}{agent} i.e. a participant on the market (an algorithm, a trader...);
    \item \textcolor{blue}{product id} i.e. a financial instrument (a share, an option...);
    \item \textcolor{blue}{direction} (buy or sell);
\end{itemize}
\end{defi}

\begin{figure}[H]
\begin{center}
\includegraphics[scale = 0.6]{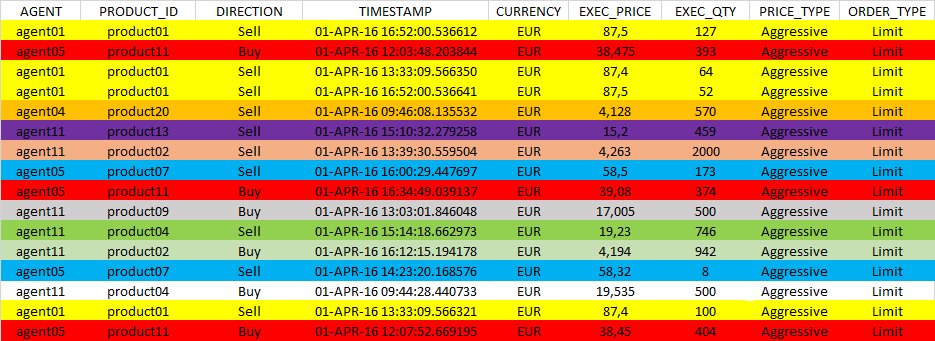}
\captionof{figure}{\textit{Sample of the history of orders \emph{before} the identification process of the algorithm. Each color corresponds to a different metaorder according to the methodology introduced in Definition \ref{defMetaorder I}.}}
\label{sample1}
\end{center}
\end{figure}

The advantage of adopting such a definition is that it is no longer necessary to work directly on raw metaorder data. Indeed, a series of orders executed by the same actor on the same product on the financial market will behave like a metaorder and therefore can be considered as such (Figure \ref{sample1}).

\begin{figure}[H]
\begin{center}
\includegraphics[scale = 0.6]{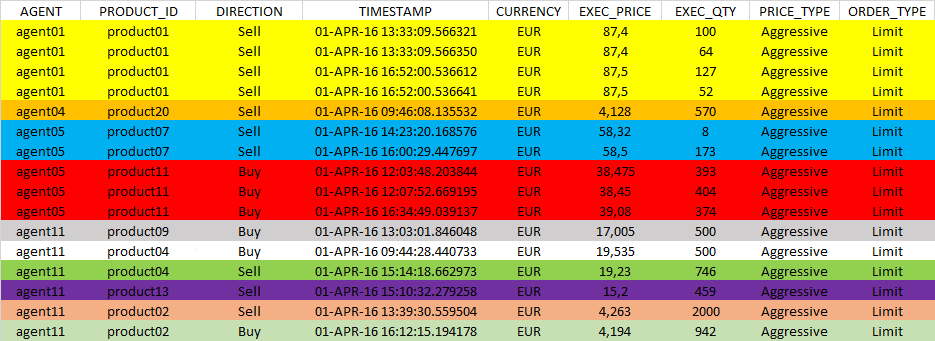}
\captionof{figure}{\textit{Sample of the history of orders \emph{after} the identification process of the algorithm. Each color corresponds to a different metaorder according to the methodology introduced in Definition \ref{defMetaorder I}.}}
\label{sample2}
\end{center}
\end{figure}

Let us mention that the current study is not restricted to metaorders executed on a single market, due to the fact that an instrument can be simultaneously traded on several markets. Also note that orders executed during the same second are aggregated in order to avoid time stamping issues: the quantities are summed up, the local VWAP is set as the execution price, and the TIMESTAMP of the last order is retained for all orders during the same second. One can see below (see Figure \ref{sample3}) how those simplifications reduce the complexity of Figure \ref{sample2}:

\begin{figure}[H]
\begin{center}
\includegraphics[scale = 0.6]{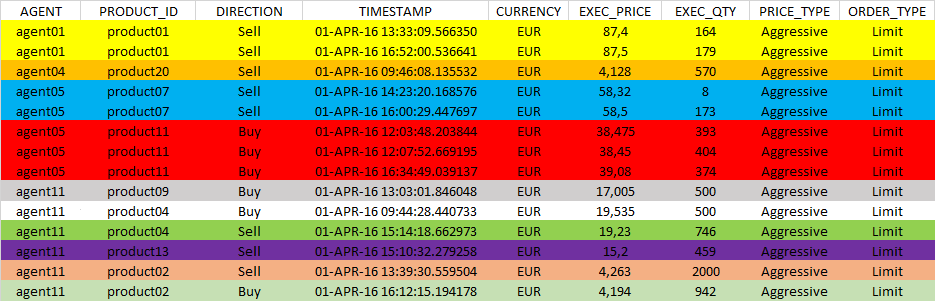}
\captionof{figure}{\textit{Sample of the history of the orders executed corresponding to the previous figure after merging the orders. The yellow metaorder which was previously a metaorder of length $N = 4$ is now a metaorder of length $N = 2$.}}
\label{sample3}
\end{center}
\end{figure}

\subsection{Algorithmic procedure}

To carry out the study of market impact, the algorithm is trained and run on a proprietary database. Figure \ref{sample1} in the previous section represents the database in its initial state, as the input of the algorithm. Figure \ref{sample2} shows an intermediate state of the data extracted from the initial data, during the metaorder reconstruction phase. Figure \ref{sample3} displays the data in their final state, directly exploitable for statistical studies. Note that the recovery and cleaning of market data are done simultaneously during the metaorder reconstruction.

\subsection{Market Impact definitions}

The same framework as that in \cite{bacry2015market} is adopted. Let $\Omega$ be the set of metaorders under scrutiny, that is, metaorders that are fully executed during a single market session, and pick $\omega \in \Omega$ executed on an instrument $S$ and during a given day $d$. Its execution starts at some time $t_0(\omega)$ and ends the same day at time $t_0(\omega) + T(\omega)$. Thus $T(\omega)$ represents the duration of the metaorder.
Denote by $Q(\omega)$ and N($\omega$) respectively the number of shares and the number of orders that have been executed during the life cycle of the metaorder $\omega$. Hence $Q(\omega)$ is the size and $N(\omega)$ the length of $\omega$. Let $V(\omega)$ be the volume traded the same day $d$ on the instrument $S$ (summed over all European trading venues) between time $t_0$ and $t_0 + T$, i.e. during the life cycle of $\omega$. The sign of $\omega$ will be denoted by $\epsilon(\omega)$ with $\epsilon = 1$ for a buy order and $\epsilon = -1$ for a sell order. Clearly, all the quantities introduced in this section depend on $\omega$. For the sake of simplicity, we chose to omit this dependence whenever there is no ambiguity and will often write $T$, $N$, $Q$, $V$, $\epsilon$ instead of $T(\omega)$, $N(\omega)$, $Q(\omega)$, $V(\omega)$, $\epsilon(\omega)$.

The \textbf{market impact curve} of a metaorder $\omega$ measures the magnitude of the relative price variation between the starting time of the metaorder $t_0$ and the current time $t > t_0$. Let $\mathcal{I}_{t}(\omega)$ be a proxy for the realized price variation between time $t_0$ and time $t_0 + t$. In line with many authors \cite{almgren2005direct} \cite{bershova2013non} \cite{bacry2015market}, we use the \textit{return proxy} defined by
\begin{equation}
\mathcal{I}_{t} = \displaystyle\frac{P_t - P_{t_0}}{P_{t_0}},
\end{equation}
where $P_t$ represents either the execution price of the financial instrument $S$ during the \textit{execution part} of the metaorder, or the mid-price during its \textit{relaxation part} starting when the metaorder has been fully executed. This estimation relies on the assumption that the \textit{exogenous market moves} $W_t$ will cancel out once averaged, i.e. as a random variable, $W_t$ should have finite variance and basically satisfy $\mathbb{E}(\epsilon(\omega)W_{t}(\omega)) = 0 $. One can thus write
\begin{equation}
\epsilon(\omega)\mathcal{I}_{t}(\omega) = \eta_{t}(\omega) + \epsilon(\omega)W_{t}(\omega),
\end{equation}
where $\eta_{t}(\omega)$ represents the market impact curve and $W_{t}(\omega)$, the exogenous variation of the price corresponding to the relative price move that would have occurred if the metaorder had not been sent to the market.

\section{Empirical study}
\label{empirical}

\subsection{Notations}

\begin{center}
\begin{tabular}{l l}
\toprule[0.15 em]
Notation & \multicolumn{1}{c}{Definition} \\
\midrule[0.1 em]
\centering $\omega$ & \centering a metaorder \tabularnewline
\centering $S(\omega)$ & \centering financial instrument of the metaorder $\omega$ \tabularnewline
\centering $t_0(\omega)$ & \centering start time of the metaorder $\omega$ \tabularnewline
\centering $T(\omega)$ & \centering duration of the metaorder $\omega$ \tabularnewline
\centering $N(\omega)$ & \centering length of the metaorder $\omega$ \tabularnewline
\centering $Q(\omega)$ & \centering size of the metaorder $\omega$ \tabularnewline
\centering $V(\omega)$ & \centering volume traded on the day $d(\omega)$ on the instrument $S(\omega)$ during $\left[t_0(\omega), t_0(\omega) + T(\omega)\right]$ \tabularnewline
\centering $\epsilon(\omega)$ & \centering sign of the metaorder $\omega$ \tabularnewline
\centering $P(\omega)$ & \centering price of $S(\omega)$ \tabularnewline
\centering $\Omega$ & \centering set of all the metaorders identified by the algorithm \tabularnewline
\centering $\Omega_{n^*} \subset \Omega $ & \centering subset of the metaorders with $N \geq n{*}$ \tabularnewline
\bottomrule[0.15 em]
\end{tabular}
\captionof{table}{\textit{Notations and definitions}}
\label{tab2}
\end{center}

\begin{rem}
As we only consider metaorders that have at least two executed transactions, $\Omega = \Omega_2$.
\end{rem}

\subsection{Aggressive Limit Orders} \label{Aggressive Limit Orders}

The subject of interest of this section is the metaorders generated by aggressive limit orders, that is, limit orders that actually cross the spread in order to trigger an immediate transaction. Such orders are sometimes rather loosely considered as market orders in the modelling literature on limit order books, but it is clear that they behave differently, as their execution price is always equal to that of the best available limit and can never trigger transactions at higher (buy) or lower (sell) price.

\subsubsection{Data}

\begin{itemize}
    \item Study period : \textbf{1st Jan 2016 - 31st Dec 2017}
    \item Markets : \textbf{European Equity Markets}
    \item Order types : \textbf{Aggressive Limit Orders}
    \item Filters : \textbf{metaorders $\omega \in \Omega$}
    \item Number of metaorders : \textbf{1 561 505}
\end{itemize}

\paragraph{Duration distribution}

\begin{figure}[H]
\begin{center}
\includegraphics[scale=0.36]{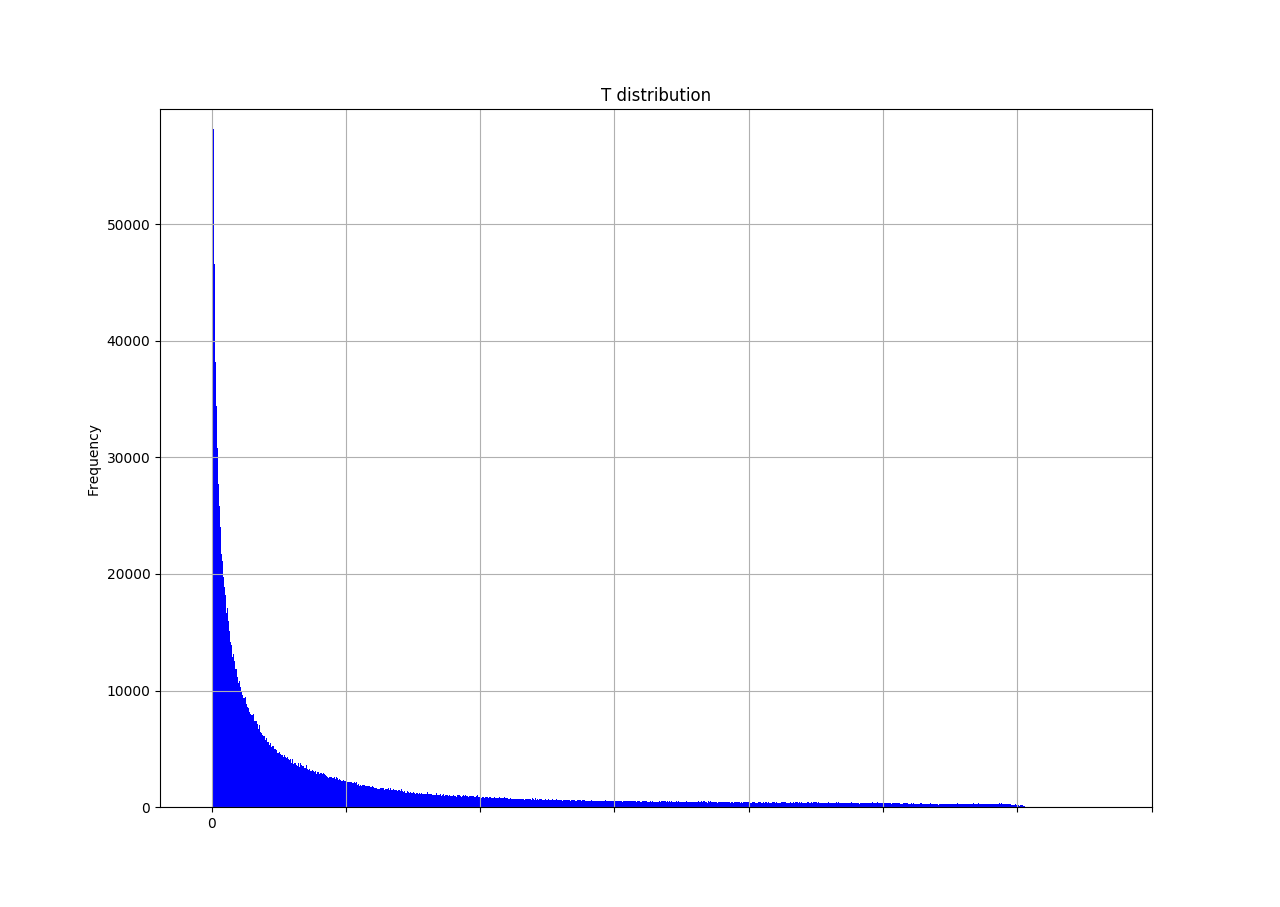}
\captionof{figure}{Duration distribution of the aggressive metaorders.}
\label{T distribution aggressive}
\end{center}
\end{figure}

One can observe that, in agreement with the intuition, metaorders with shorter durations are more frequent (Figure \ref{T distribution aggressive}).

\paragraph{Length distribution}

\begin{figure}[H]
\begin{center}
\includegraphics[scale=0.36]{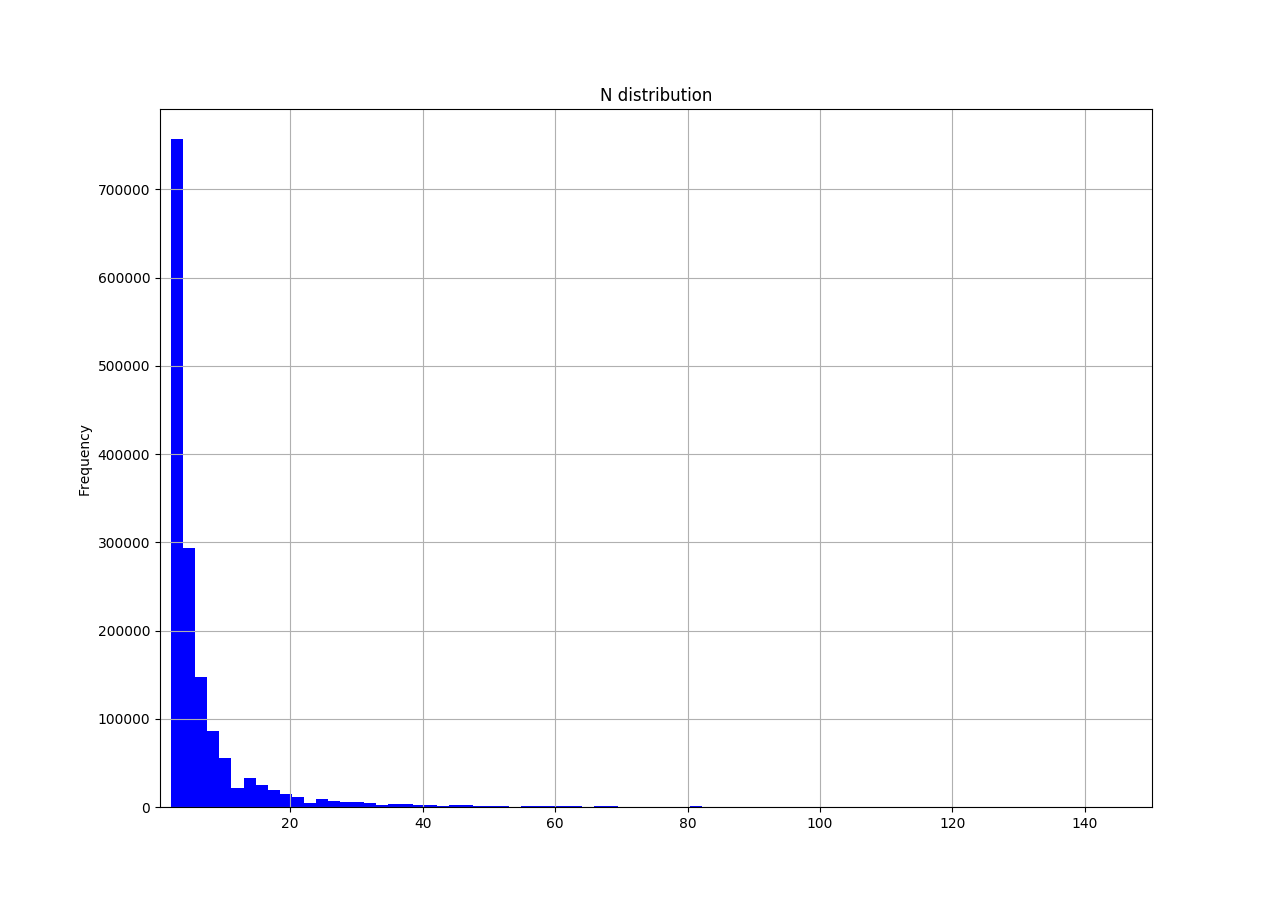}
\captionof{figure}{Length distribution of the aggressive metaorders (mean: 8, median: 5).}
\label{N distribution aggressive}
\end{center}
\end{figure}

\begin{figure}[H]
\begin{center}
\includegraphics[scale=0.36]{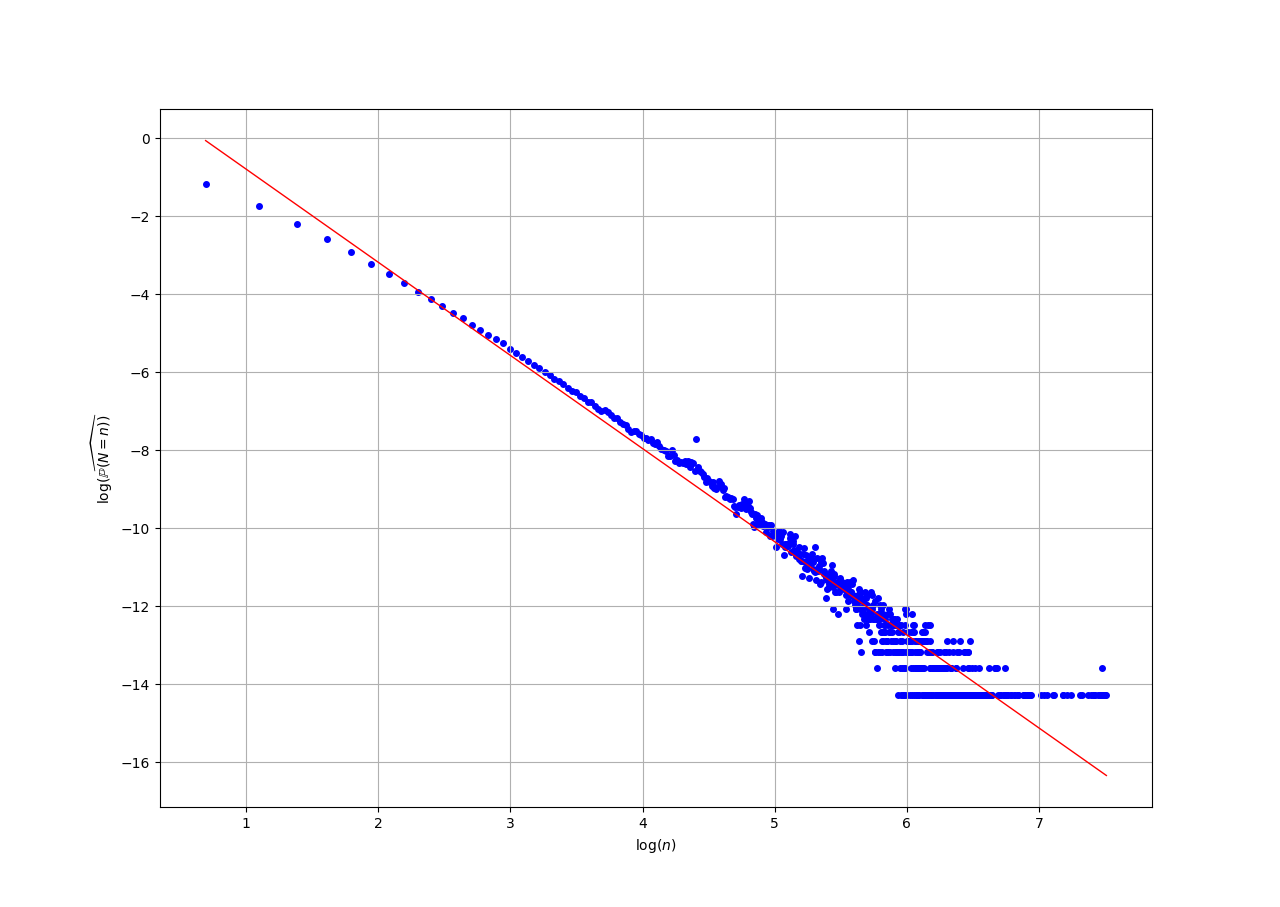}
\captionof{figure}{Fitting Length distribution of the aggressive metaorders.} 
\label{N distribution aggressive loglog}
\end{center}
\end{figure}

As already observed for the duration, shorter metaorders (in length) are more represented. This observation is obviously not a surprise, since the quantities $N$ and $T$ are expected to be highly positively correlated.  A log-log scale (Figure \ref{N distribution aggressive loglog}) gives a more precise idea of the distribution of $N$. The apparently linear relation suggests that $N$ follows a discrete Pareto distribution.

$\widehat{\mathbb{P}(N = n)}$ stands for the natural frequency estimator of the probability $\mathbb{P}(N = n)$.
As regards the parameter $\beta$ introduced in \cite{farmer2013efficiency} - see Section \ref{Dependence on the metaorder size distribution} - one obtains the estimate $\beta \approx 1.4$. Therefore, $N$ is distributed as a power law, in agreement with \cite{vaglica2008scaling} who reconstructed metaorders on the Spanish stock exchange using data with brokerage codes and found that $N$ is distributed as power law for large $N$ with $\beta \approx 1.7$. 

\paragraph{Participation rate distribution}

\begin{figure}[H]
\begin{center}
\includegraphics[scale=0.36]{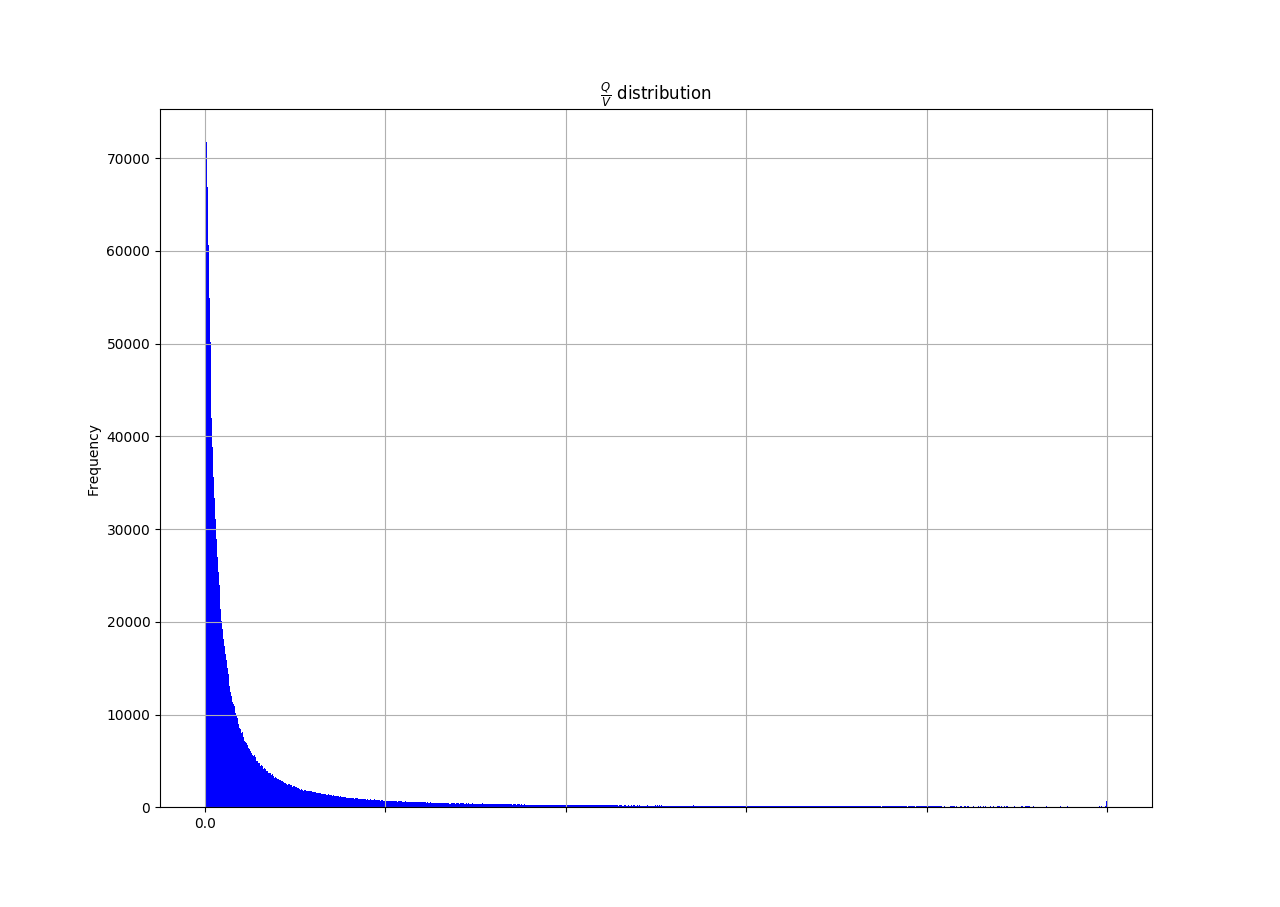}
\captionof{figure}{$\displaystyle\frac{Q}{V}$ distribution of the aggressive metaorders.}
\label{Q/V distribution aggressive}
\end{center}
\end{figure}

Similar to the observations for duration and length (see Figures \ref{T distribution aggressive} and \ref{N distribution aggressive}), metaorders with smaller participation rates are more represented. Again, and as expected, the quantities $N$ and $\displaystyle\frac{Q}{V}$ are highly positively correlated (Figure \ref{Q/V distribution aggressive}).

\subsubsection{Market Impact curves}

The main results of Section \ref{Aggressive Limit Orders} are now given, namely, the \textit{market impact curves} for aggressive metaorders. In order to plot the market impact dynamics, a \textit{bucketing method} is used: Consider for example that one wants to plot $y$ as a function of $x$, $x, y$ being two arrays of data. One starts by ordering the couple of values $(x_{i},y_{i})$ according to the values of $x$ and then divides the \textbf{sorted (by $x$) distribution $(x,y)_{sorted}$} into $N_{bucket}$. This procedure yields $N_{bucket}$ subsets of the distribution $(x,y)_{sorted}$, $(x_{i},y_{i})_{i \in I_{1}}$, $(x_{i},y_{i})_{i \in I_{2}}$, ..., $(x_{i},y_{i})_{i \in I_{N_{bucket}}}$, and for each bucket $I_k$ the \textit{mean values} $(\overline{x}_{k}, \overline{y}_{k})$ is computed. The last step of this bucketing method is to plot the points $(\overline{x}_{1}, \overline{y}_{1})$, $(\overline{x}_{2}, \overline{y}_{2})$, ..., $(\overline{x}_{N_{bucket}}, \overline{y}_{N_{bucket}})$.

\paragraph{Market Impact Dynamics} \label{Market Impact Dynamics Aggressive}

To study the dynamics of the market impact, one plots $(\epsilon(\omega)\mathcal{I}_t(\omega))_{\omega \in \Omega, t_0(\omega) \leq\, t \,\leq t_0(\omega) + 2T(\omega)}$. The first sub-interval $t_0(\omega) \leq\, t \,\leq t_0(\omega) + T(\omega)$ corresponds to the execution of the metaorder, whereas the second $t_0(\omega) + T(\omega)  \leq\, t \,\leq t_0(\omega) + 2T(\omega)$ corresponds to the relaxation. The study of relaxation presents a degree of arbitrariness, since a choice has to be made as to the elapsed time after the metaorder is completed. For the sake of homogeneity, the relaxation is measured over the same duration as the execution. This choice seems to be a good compromise to cope with two antagonistic requirements, one being to minimize this elapsed time because of the diffusive nature of prices, the other being to maximize it so as to make sure that the relaxation is achieved. Further studies performed on our database actually show very little dependence of the permanent impact level on this time parameter, a result which we found quite comforting, and in line with some previous results in the literature, see e.g. \cite{gomes2015market}.

In order to perform an extensive statistical analysis involving metaorders of varying lengths in physical and volume time, a rescaling in time is necessary, see e.g. \cite{bacry2015market}. With this convention, all orders are executed on the time interval $[0,1]$ and price relaxation occurs in the time interval  $[1,2]$. For each metaorder $\omega$, one considers $[0,1]$ instead of $[t_0(\omega), t_0(\omega) + T(\omega)]$ $\left( [0,1] = \displaystyle\frac{[t_0(\omega), t_0(\omega) + T(\omega)] - t_0(\omega) }{T(\omega)} \right)$ for the \textit{execution part} of $\omega$ and $[1,2]$ instead of $[t_0(\omega) + T(\omega), t_0(\omega) + 2T(\omega)]$ for the \textit{relaxation part} of $\omega$, and then averages using the \textit{bucketing method} previously described on the time-rescaled volume quantities.

\begin{figure}[H]
\begin{center}
\includegraphics[scale=0.42]{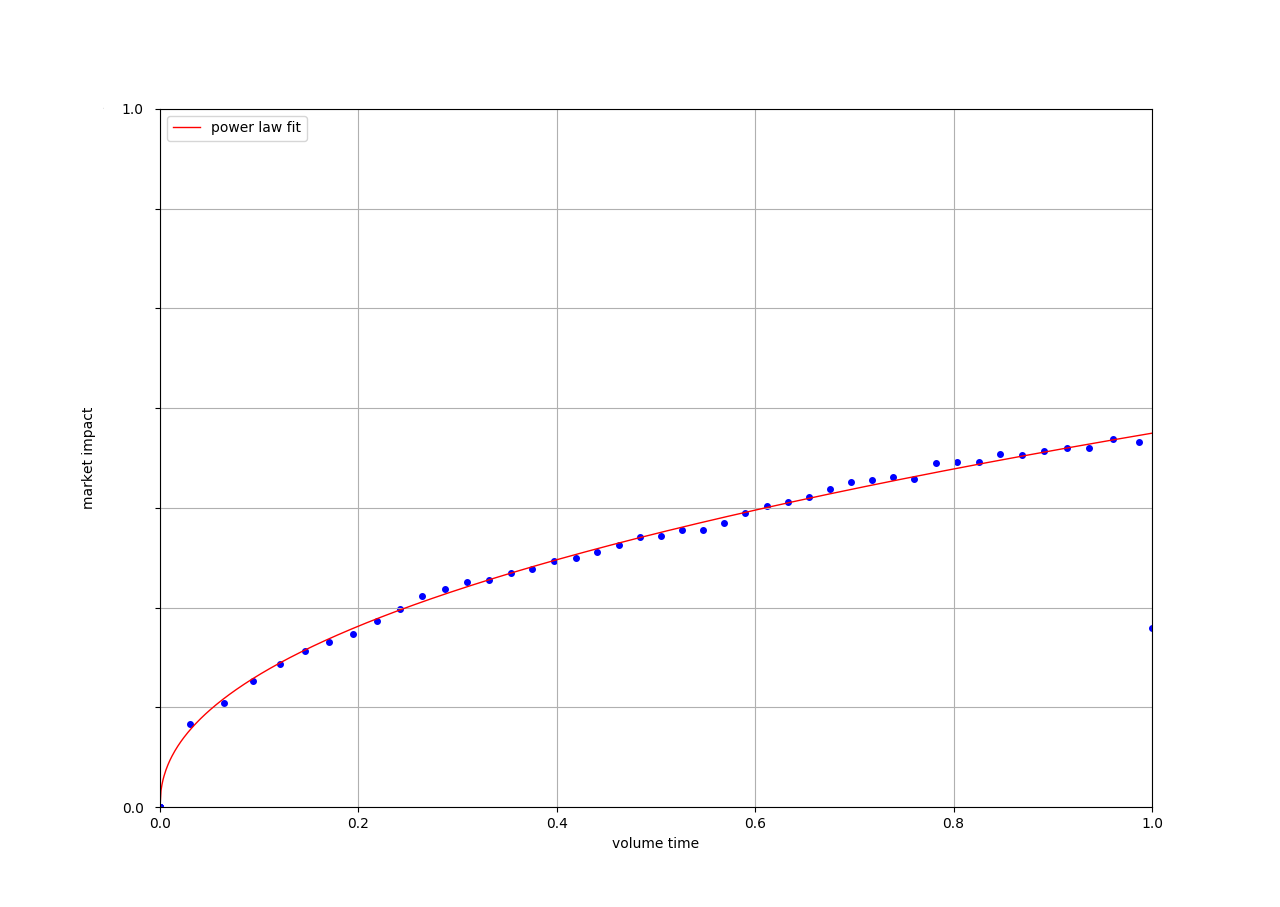}
\captionof{figure}{Market impact dynamics during the \textit{execution part} in the case of the aggressive metaorders (set: $\Omega$, 1 561 505 metaorders, power law fit: $y = 0.54 \times x^{0.45}$).}
\label{dynamics aggressive}
\end{center}
\end{figure}

The time variable $t \in [0,1]$ in Figure \ref{dynamics aggressive} is actually the \textit{volume time}, i.e., the ratio between the part of the metaorder already executed at the time of the observation and the total size of the metaorder - of course, at the end of the execution part this quantity is always equal to 1.

\begin{figure}[H]
\begin{center}
\includegraphics[scale=0.40]{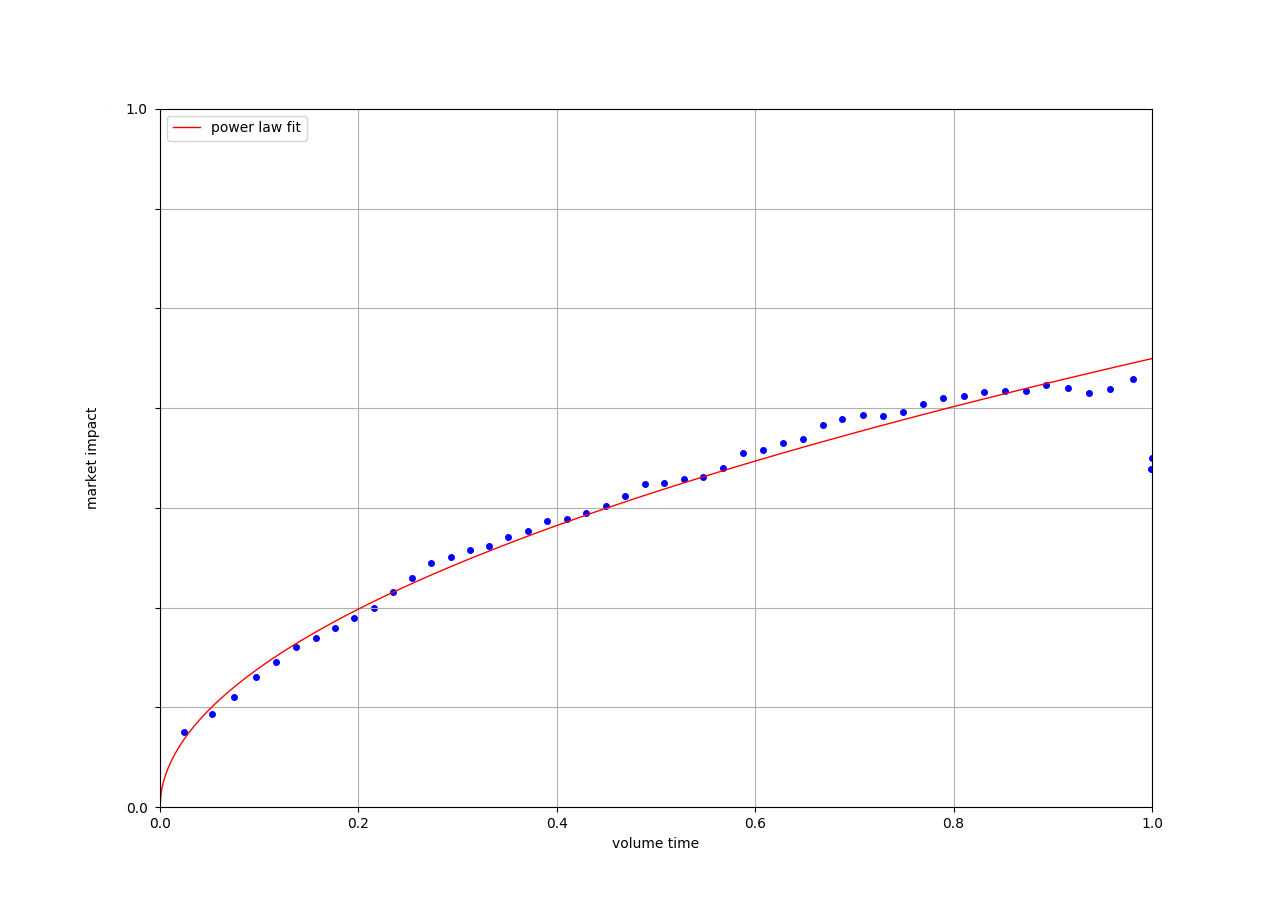}
\captionof{figure}{Market impact dynamics during the \textit{execution part} in the case of the aggressive metaorders (set: $\Omega_{10}$, 275 969 metaorders, power law fit: $y = 0.63 \times x^{0.51}$).}
\label{dynamics aggressive 10}
\end{center}
\end{figure}

\begin{figure}[H]
\begin{center}
\includegraphics[scale=0.40]{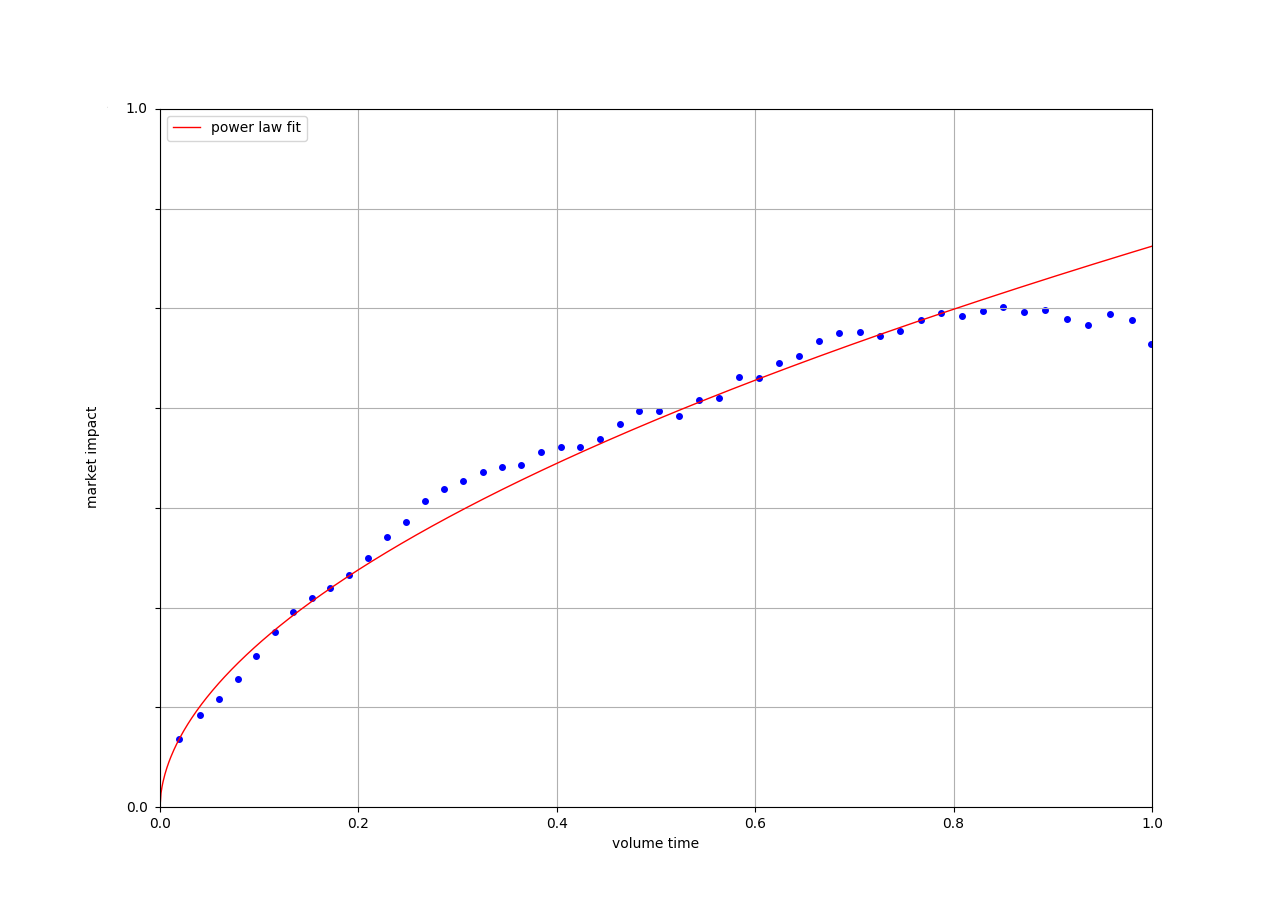}
\captionof{figure}{Market impact dynamics during the \textit{execution part} in the case of the aggressive metaorders (set: $\Omega_{30}$, 65 683 metaorders, power law fit: $y = 0.80 \times x^{0.53}$).}
\label{dynamics aggressive 30}
\end{center}
\end{figure}

\newpage
The analysis clearly yields an increasing, concave market impact curve. The decay observed in the last points of the curve is an artifact that can be explained by the larger number of metaorders of smaller lengths and with lower impact, as shown in Figure \ref{N distribution aggressive}. Also note that on the three figures \ref{dynamics aggressive}, \ref{dynamics aggressive 10} and \ref{dynamics aggressive 30}, the larger the metaorders, the higher the impacts: $0.53$, $0.60$ and then $0.71$ for the temporary market impact. However, the last curve \ref{dynamics aggressive 30} indicates that, for large metaorders, the impact reaches a plateau, as if the market has adjusted to the information it received in such a way that the execution of the metaorder no longer affects it.

\begin{figure}[H]
\begin{center}
\includegraphics[scale=0.46]{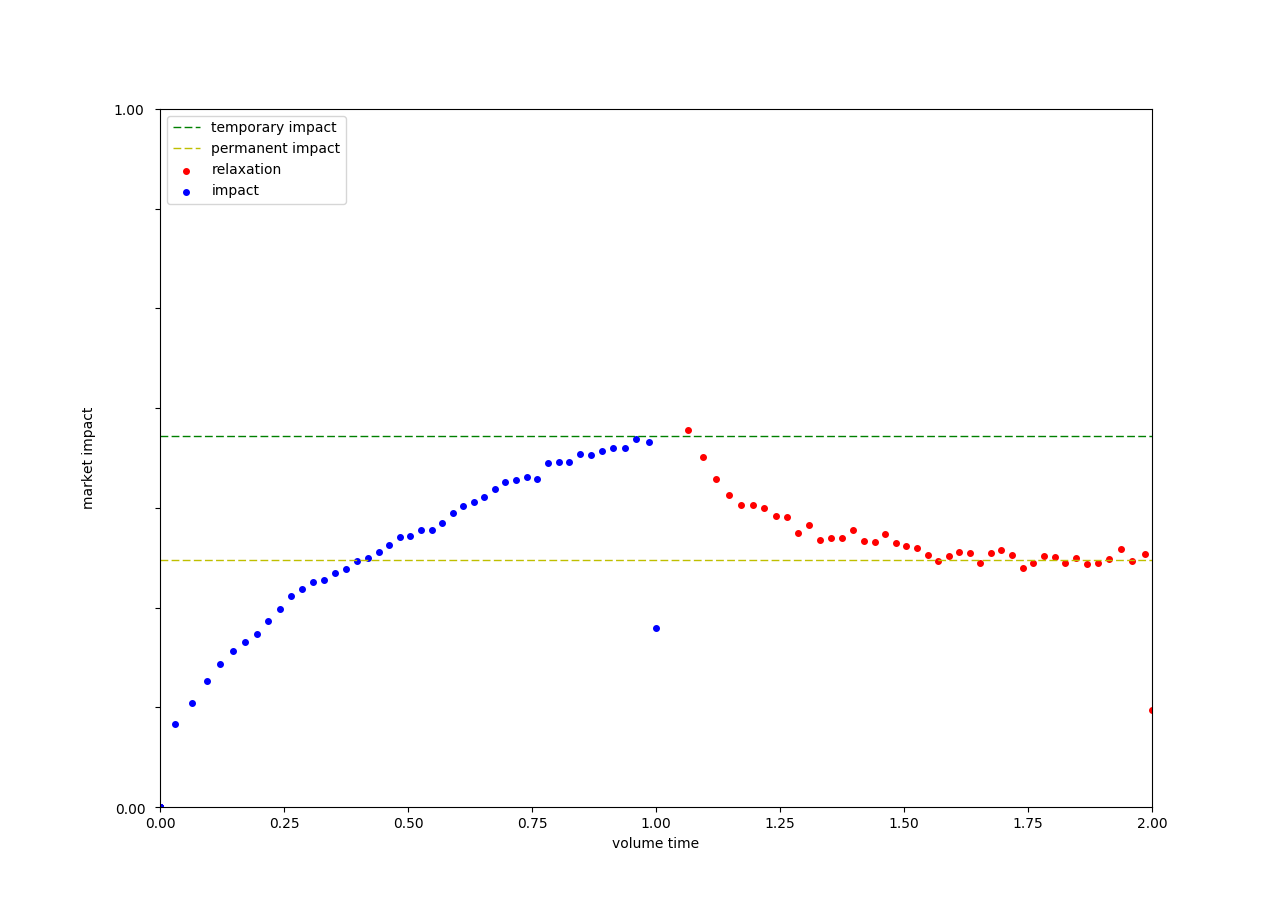}
\captionof{figure}{Market impact dynamics with relaxation in the case of the aggressive metaorders (set: $\Omega$, 1 561 505 metaorders, temporary impact: 0.53, permanent impact: 0.35).}
\label{dynamics relaxation aggressive}
\end{center}
\end{figure}

\begin{figure}[H]
\begin{center}
\includegraphics[scale=0.40]{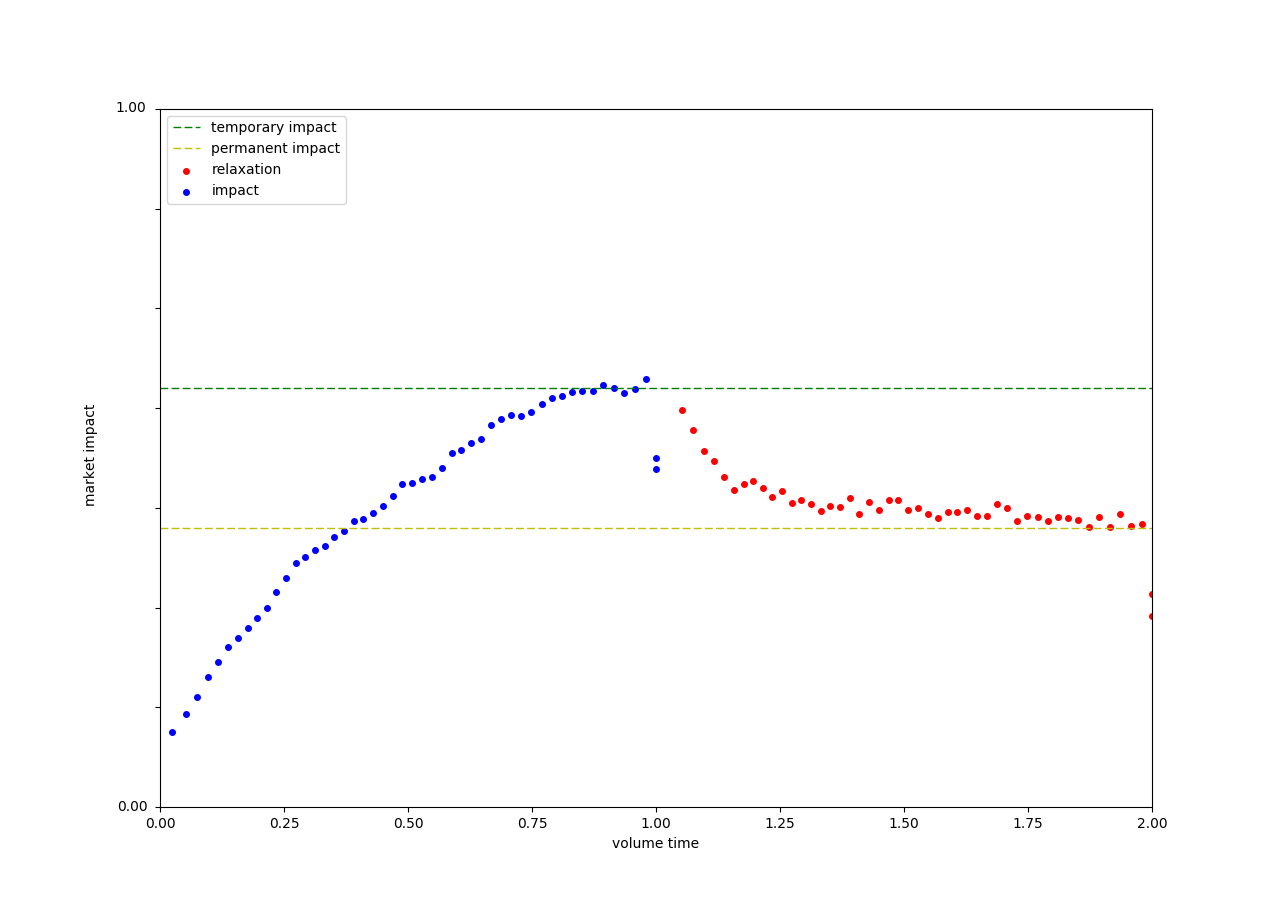}
\captionof{figure}{Market impact dynamics with relaxation in the case of the aggressive metaorders (set: $\Omega_{10}$, 275 969 metaorders, temporary impact: 0.60, permanent impact: 0.40).}
\label{dynamics relaxation aggressive 10}
\end{center}
\end{figure}

\begin{figure}[H]
\begin{center}
\includegraphics[scale=0.40]{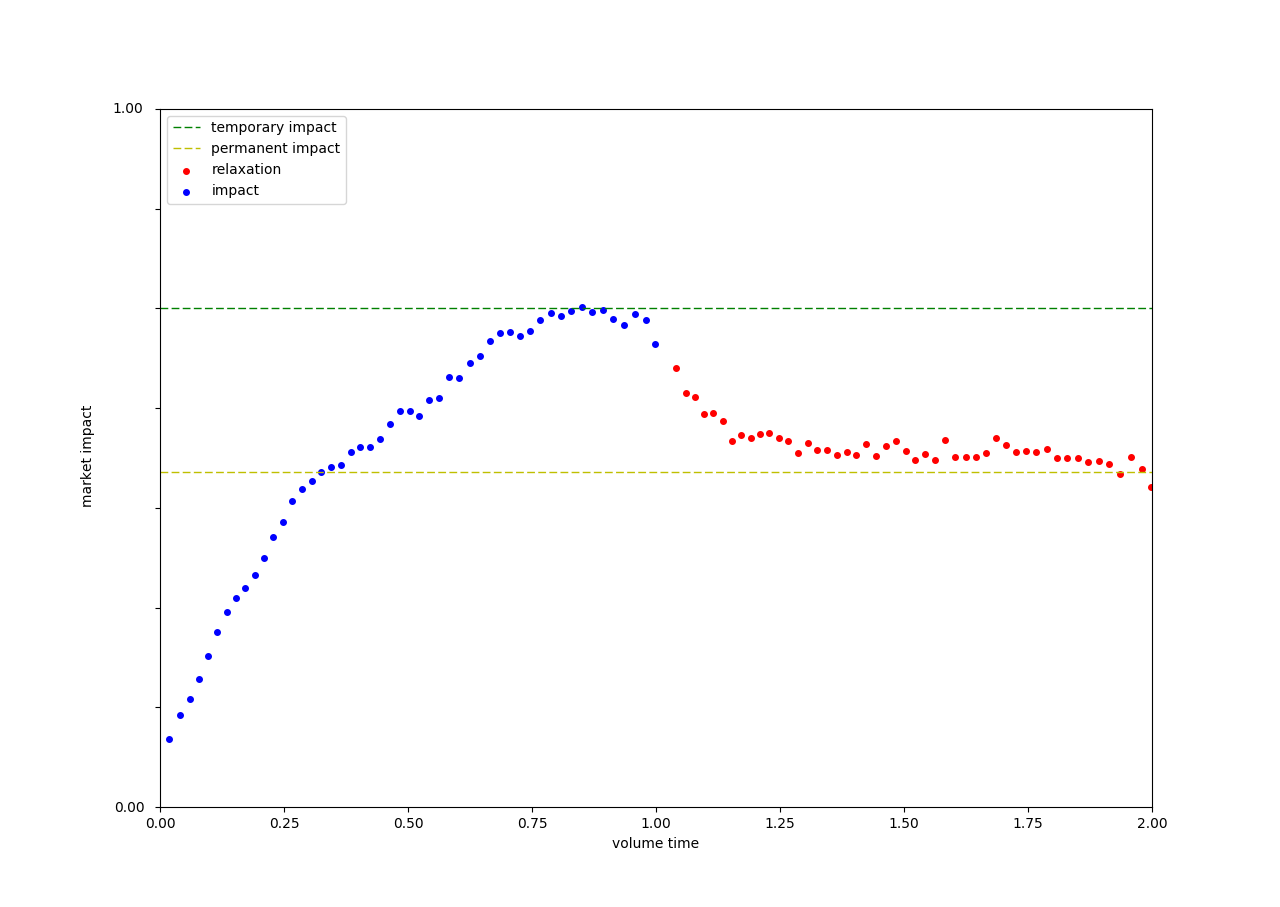}
\captionof{figure}{Market impact dynamics with relaxation in the case of the aggressive metaorders (set: $\Omega_{30}$, \,\,\,65 683 metaorders, temporary impact: 0.71, permanent impact: 0.48).}
\label{dynamics relaxation aggressive 30}
\end{center}
\end{figure}

The blue points correspond to execution prices and the red points correspond to mid-prices observed at identical times starting from the end of the metaorder. The isolated points observed in Figures \ref{dynamics relaxation aggressive} and \ref{dynamics relaxation aggressive 10} are due to the fact that we have considered all metaorders, so metaorders with a small length, especially those with length $N = 2$ are over-represented in \textbf{volume time} = 1.0, therefore inducing a bias towards the end of the curve. One can make this artifact vanish when considering only metaorders with larger sizes, see e.g. \ref{dynamics relaxation aggressive 30}.

Figures \ref{dynamics relaxation aggressive}, \ref{dynamics relaxation aggressive 10} and \ref{dynamics relaxation aggressive 30} clearly exhibit the concave shape of market impact during the execution part, followed by a convex and decreasing relaxation. Simply by eyeballing Figure \ref{dynamics relaxation aggressive}, one can safely assume that relaxation is complete and stable at a level around $0.35$. However, on Figures \ref{dynamics relaxation aggressive 10} and \ref{dynamics relaxation aggressive 30}, relaxation does not seem to be quite complete. This behaviour for larger metaorders is quite probably  due to the fact that it is not always possible to reach the final time $t_0 + 2 T$ during the intraday observation period. Hence, the relaxation of larger metaorders may be hindered by the closing of the market.

A conclusion to this section is that the concave shape of the temporary impact and the convex relaxation curve are in line with the empirical results in \cite{bacry2015market} and \cite{bershova2013non}. Also, and more interestingly, the market impact and relaxation curves confirm the theoretical findings of \cite{farmer2013efficiency} that the impact should be concave and increasing, and that the final impact after the execution is performed should relax to about two-thirds of the peak impact.

\subsection{Execution strategies}

The purpose of this section is to analyse metaorders executed \emph{via} passive limit orders, and compare the resulting market impact curves with those previously obtained. Note that the scales on the graphs concerning the aggressive and passive metaorders are exactly the same, thereby making comparisons possible.

At this stage, some comments are in order: by nature, passive limit orders are not always executed and therefore, the notion of market impact for such orders has to be taken with a grain of salt. When an agent places a new passive limit order, especially within the bulk of the order book, it is not clear whether this should indicate that the market is moving in any direction at all, and an impact should not always be expected. In fact, a strategy purely based on limit orders can be even considered to have only negative market impact, since the order is executed only after the price has moved in favour of the agent. And of course, an execution strategy relying only on limit orders will often fail to achieve its target, thereby facing an implementation shortfall that will have to be dealt at some point in the future.

In order to cope with this inherent difficulty  and also to present results that are consistent with the task at hand, namely the study of market impact during the execution of a metaorder, the analyses in this section are performed on a database of \textit{execution strategies}. Metaorders in this database are executed using (on average) $\sim 65\%$ of passive limit orders and $\sim 30\%$ of aggressive limit orders, the remaining $5\%$ consisting of orders of various types (market, market on close...). 

\subsubsection{Data}

\begin{itemize}
    \item Study period : \textbf{1st Jan 2016 - 31st Dec 2017}
    \item Markets : \textbf{European Equity Markets}
    \item Order types : \textbf{All Orders}\footnote{$\sim 65\%$ passive limit orders, $\sim 30\%$ aggressive limit orders and $5\%$ others}
    \item Filters : \textbf{metaorders $\omega \in \Omega$}
    \item Number of metaorders : \textbf{74 552}
\end{itemize}

\paragraph{Duration distribution}

\begin{figure}[H]
\begin{center}
\includegraphics[scale=0.33]{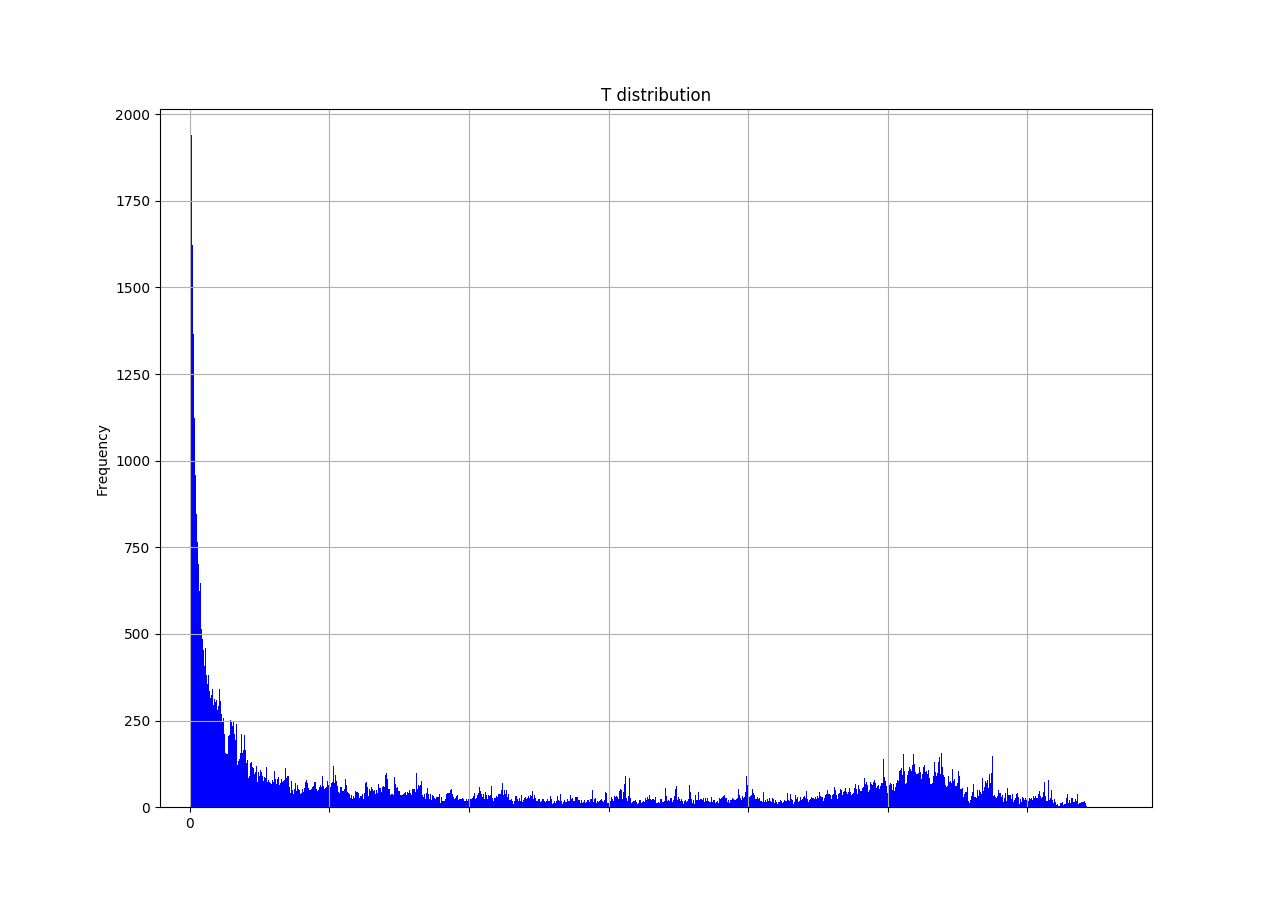}
\captionof{figure}{Duration distribution in seconds of the execution metaorders.}
\label{T distribution execution}
\end{center}
\end{figure}

As expected, and previously shown for the aggressive metaorders, see Figure \ref{T distribution aggressive}, one observes again (Figure \ref{T distribution execution}) that, the shorter the metaorders are, the more they are represented. Different from the case of the aggressive metaorders, the distribution is seen to have a fatter tail, an indication of the "intelligence" of the execution strategies.

\paragraph{Length distribution}

\begin{figure}[H]
\begin{center}
\includegraphics[scale=0.33]{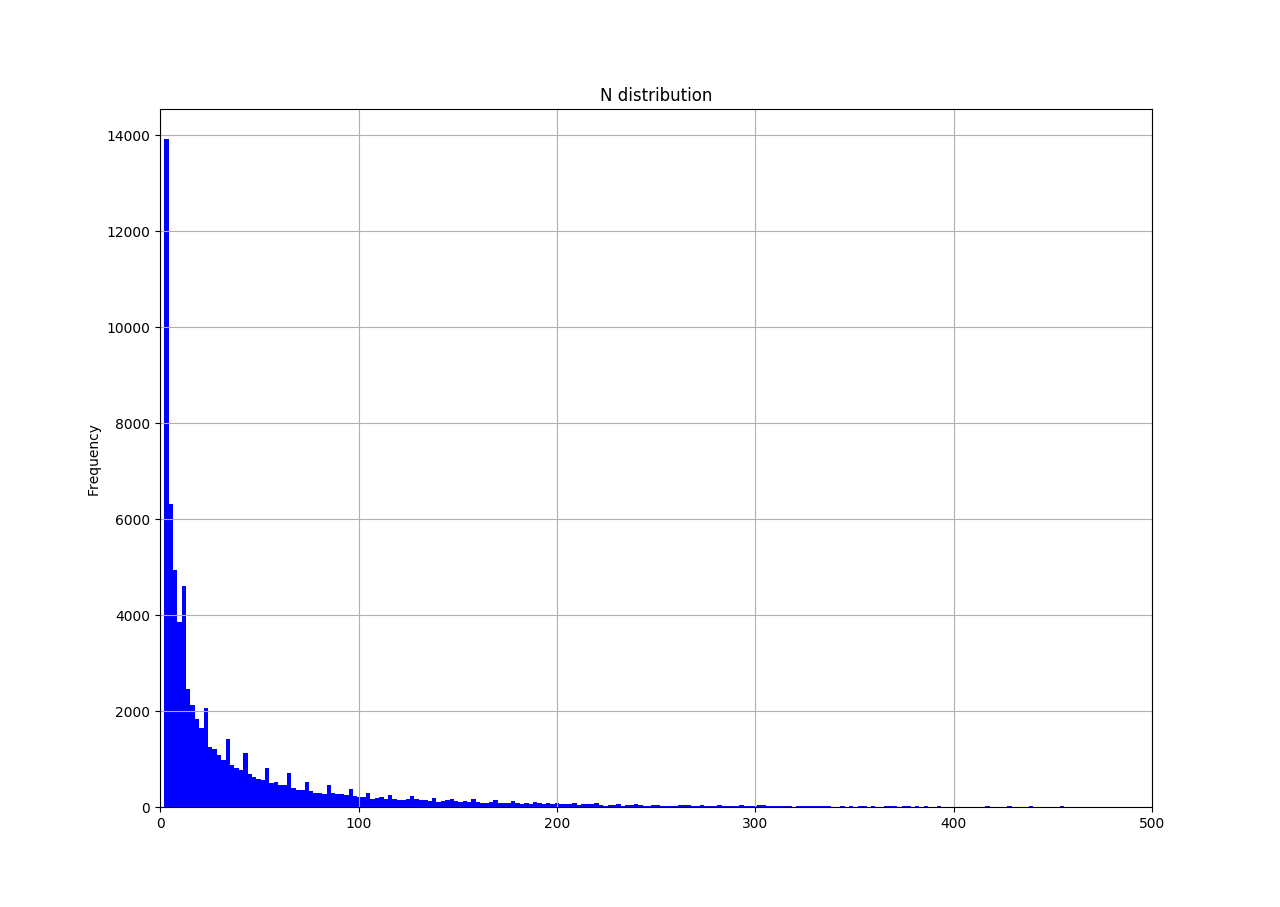}
\captionof{figure}{Length distribution of the execution metaorders (mean: 50, median: 17).}
\label{N distribution execution}
\end{center}
\end{figure}

The shape of the distribution in Figure \ref{N distribution execution} is very similar to that obtained for the aggressive metaorders in Figure \ref{N distribution aggressive}.

\begin{figure}[H]
\begin{center}
\includegraphics[scale=0.33]{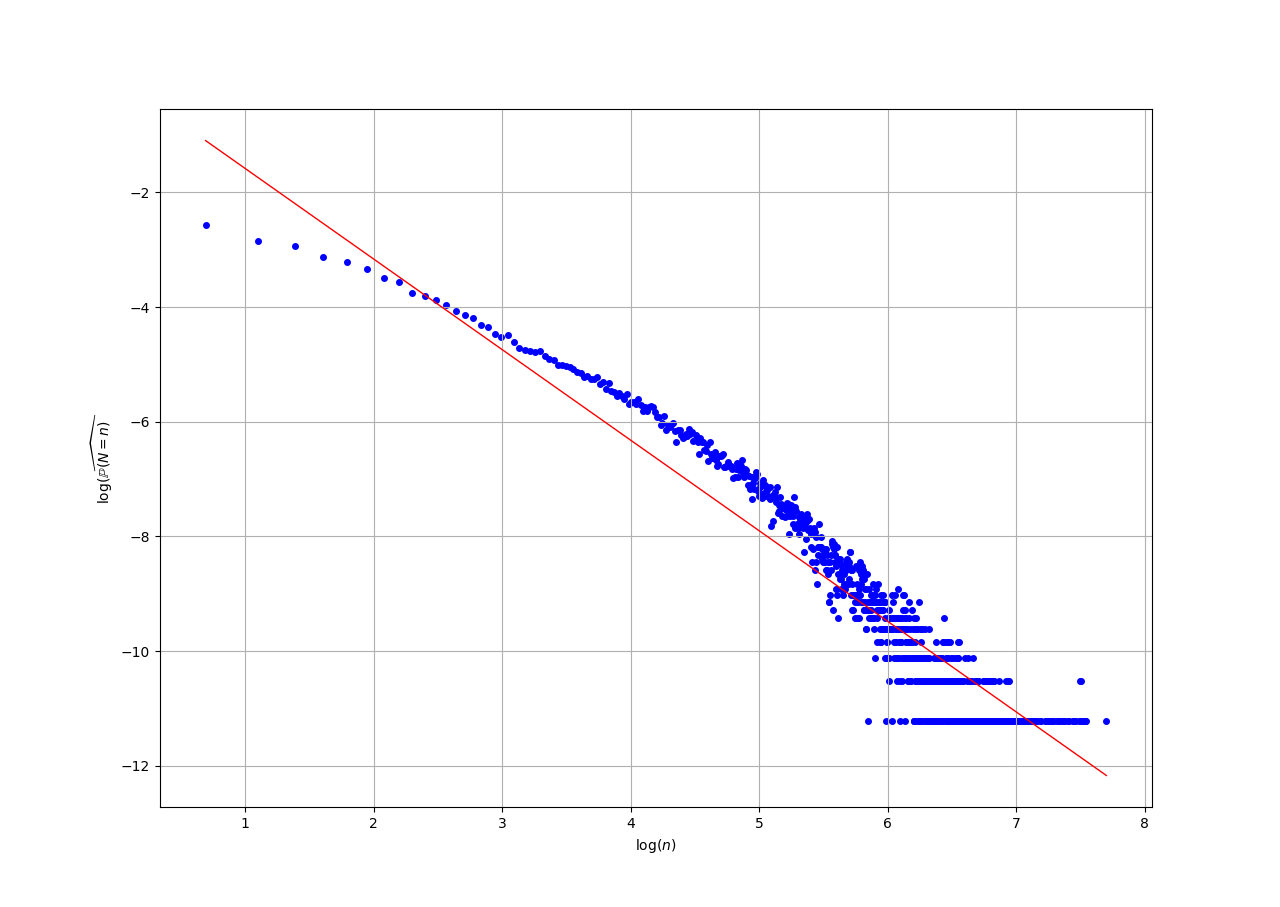}
\captionof{figure}{Length distribution of the execution metaorders with a log-log scale.}
\label{N distribution execution loglog}
\end{center}
\end{figure}

Again, a linear relation in log-log scale seems rather clear (Figure \ref{N distribution execution loglog}), although it is noisier, with $\beta \approx 1.8$ in this case. This tends to confirm the relevance of the discrete Pareto distribution independently of the nature of the orders.

\paragraph{Participation rate distribution}

\begin{center}
\includegraphics[scale=0.33]{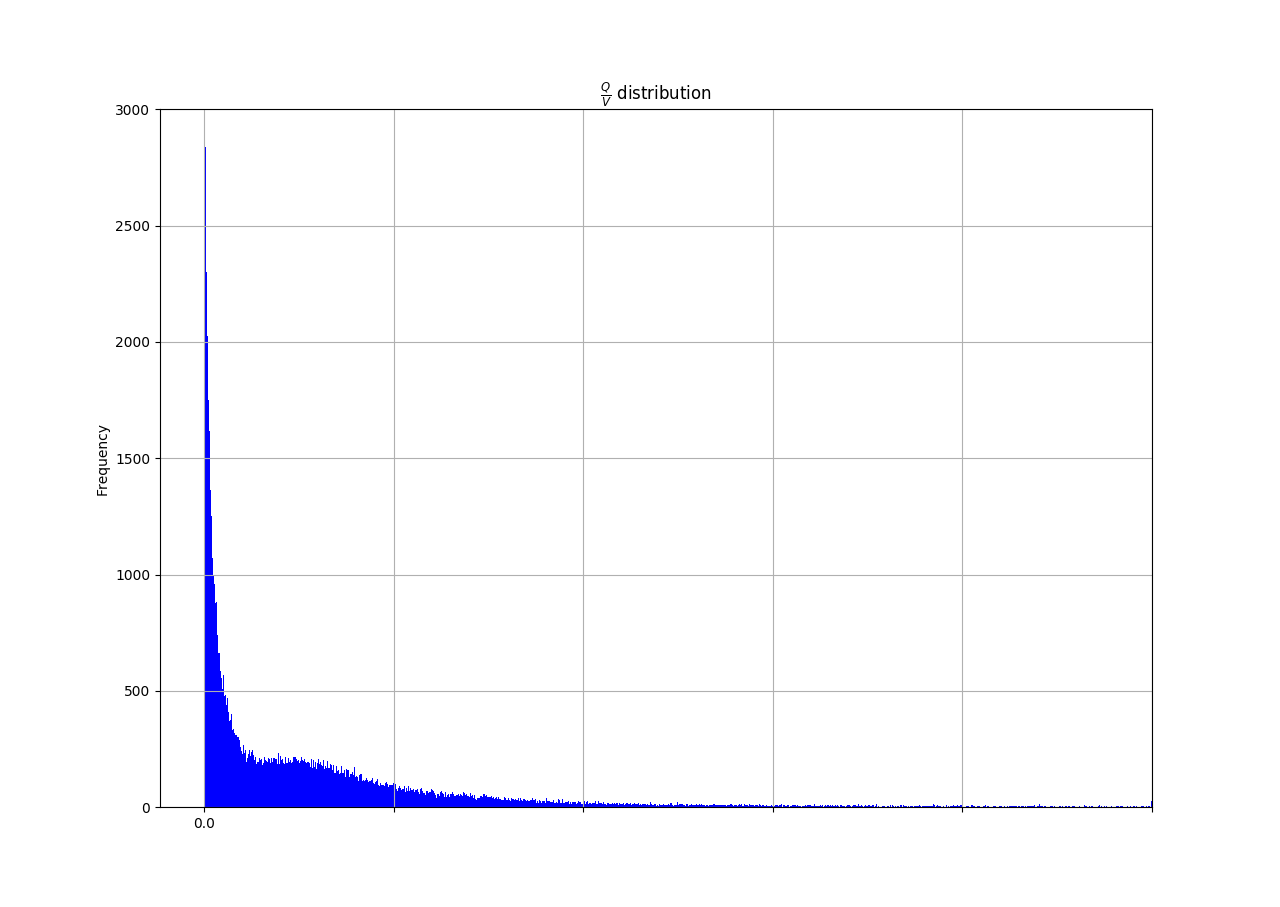}
\captionof{figure}{$\displaystyle\frac{Q}{V}$ distribution of the execution metaorders.}
\label{Q/V distribution execution}
\end{center}

Once again the quantities $N$, $T$ and $\displaystyle\frac{Q}{V}$ are highly positively correlated (Figure \ref{Q/V distribution execution}). Besides, and not surprisingly, the execution metaorders have larger participation rates than those observed in the aggressive case (Figure \ref{Q/V distribution aggressive}).

\subsubsection{Market Impact Dynamics}

The market impact study proceeds along the same lines as those presented in \ref{Market Impact Dynamics Aggressive}. Hence, to plot $(\epsilon(\omega)\mathcal{I}_t(\omega))_{\omega \in \Omega, t_0(\omega) \leq\, t \,\leq t_0(\omega) + T(\omega)}$, one has to consider its rescaled time version $(\epsilon(\omega)\mathcal{I}_s(\omega))_{\omega \in \Omega, 0 \leq\, s \,\leq 1}$, i.e. for each metaorder $\omega$ we consider $[0,1]$ instead of $[t_0(\omega), t_0(\omega) + T(\omega)]$ such as $ [0,1] = \displaystyle\frac{[t_0(\omega), t_0(\omega) + T(\omega)] - t_0(\omega) }{T(\omega)} $.

\begin{figure}[H]
\begin{center}
\includegraphics[scale=0.46]{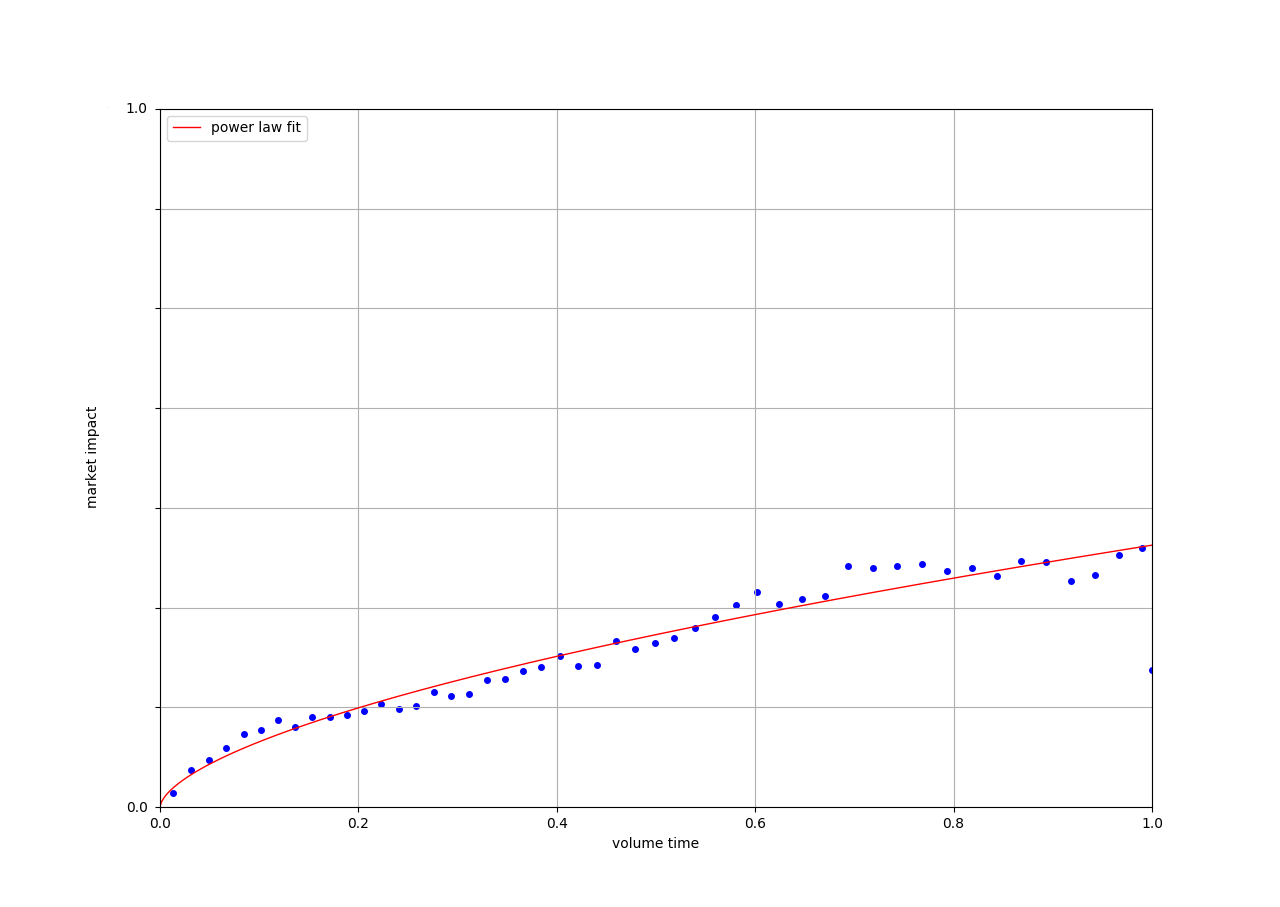}
\captionof{figure}{Market impact dynamics during the \textit{execution part} in the case of the execution metaorders (set: $\Omega$, 74 552 metaorders, fit: $y = 0.37 \times x^{0.60}$).}
\label{dynamics execution}
\end{center}
\end{figure}

\begin{figure}[H]
\begin{center}
\includegraphics[scale=0.40]{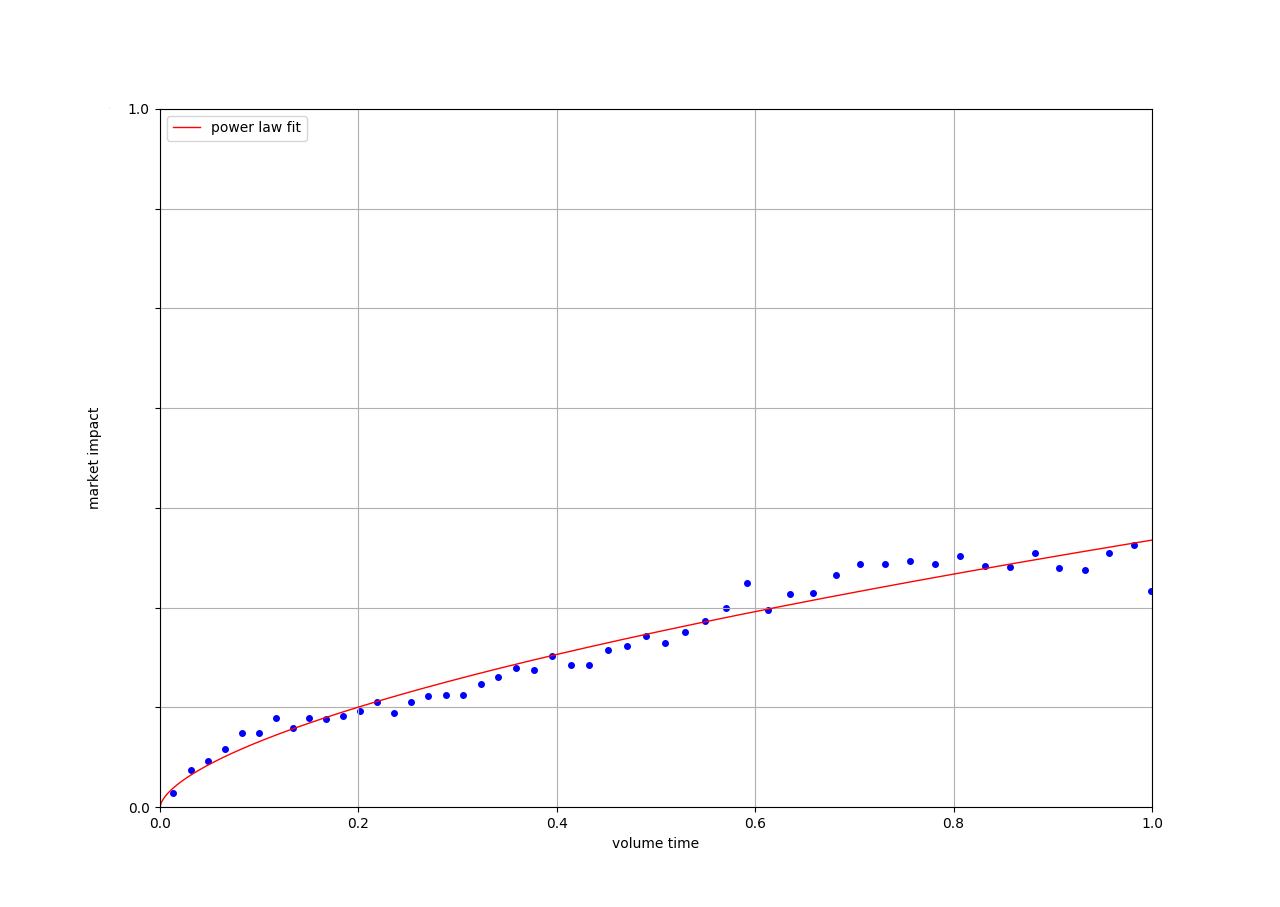}
\captionof{figure}{Market impact dynamics during the \textit{execution part} in the case of the execution metaorders (set: $\Omega_{10}$, 47 243 metaorders, power law fit: $y = 0.37 \times x^{0.61}$).}
\label{dynamics execution 10}
\end{center}
\end{figure}

\begin{figure}[H]
\begin{center}
\includegraphics[scale=0.40]{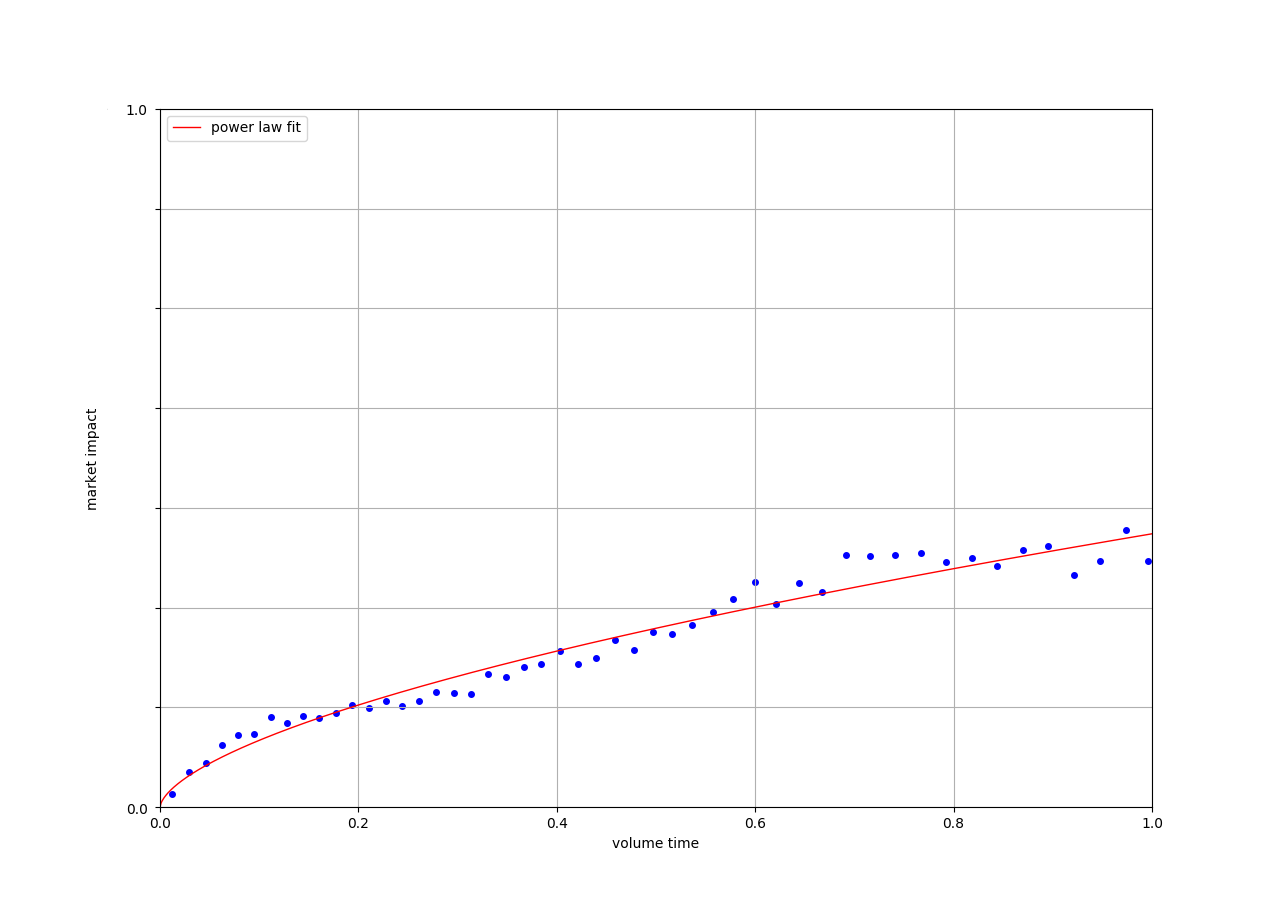}
\captionof{figure}{Market impact dynamics during the \textit{execution part} in the case of the execution metaorders (set: $\Omega_{30}$, 27 710 metaorders, power law fit: $y = 0.40 \times x^{0.62}$).}
\label{dynamics execution 30}
\end{center}
\end{figure}

One observes a positive, increasing and concave market impact curve. The impacts appear to be smaller than in the aggressive case (Figure \ref{dynamics aggressive}) although participation rates are higher and metaorders last longer. Moreover, the three curves Figures \ref{dynamics execution}, \ref{dynamics execution 10} and \ref{dynamics execution 30} show that the impact is insensitive to the size of the metaorders. These two features are a clear indication that it is beneficial to use passive limit orders. An execution metaorder has the main constraint of executing on the market a known and predefined size in advance, whatever the chosen strategy is. The choice of an execution strategy less brutal than one relying solely on aggressive metaorders allows to limit and control its impact.

\begin{figure}[H]
\begin{center}
\includegraphics[scale=0.46]{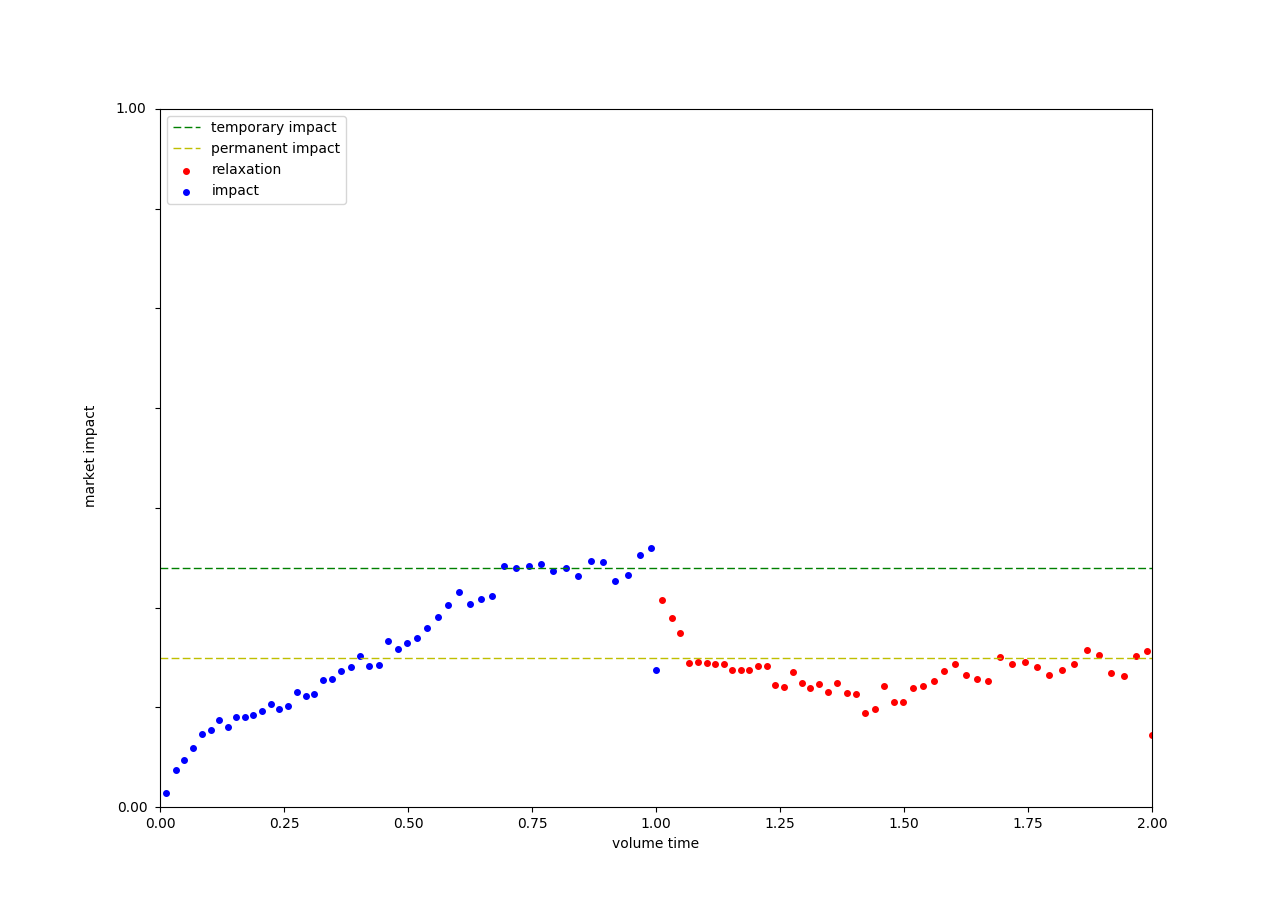}
\captionof{figure}{Market impact dynamics with relaxation in the case of the execution metaorders (set: $\Omega$, 74 552 metaorders, temporary impact: 0.34, permanent impact: 0.22).}
\label{dynamics relaxation execution}
\end{center}
\end{figure}

\begin{figure}[H]
\begin{center}
\includegraphics[scale=0.40]{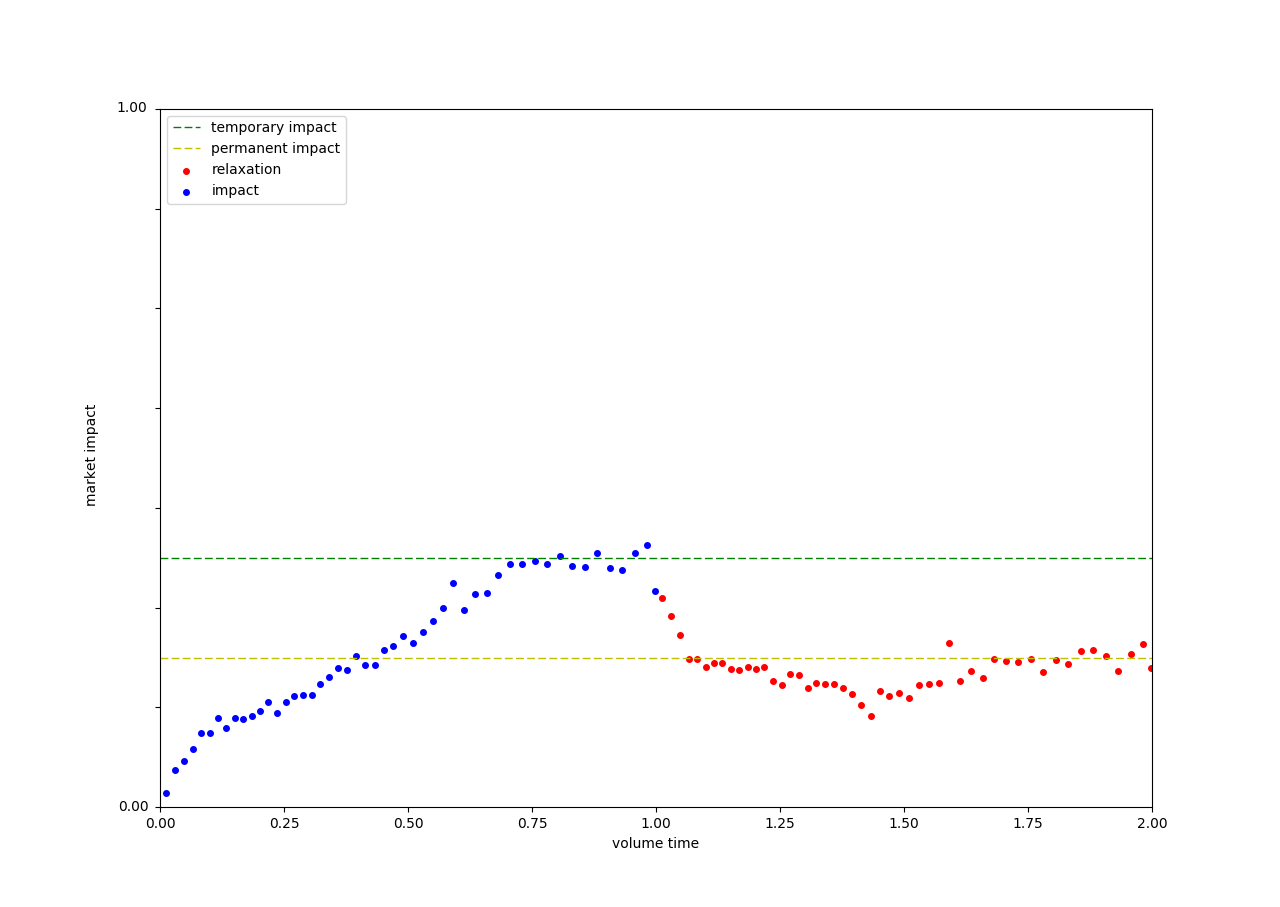}
\captionof{figure}{Market impact dynamics with relaxation in the case of the execution metaorders (set: $\Omega_{10}$, 47 423 metaorders, temporary impact: 0.36, permanent impact: 0.23).}
\label{dynamics relaxation execution 10}
\end{center}
\end{figure}

\begin{figure}[H]
\begin{center}
\includegraphics[scale=0.40]{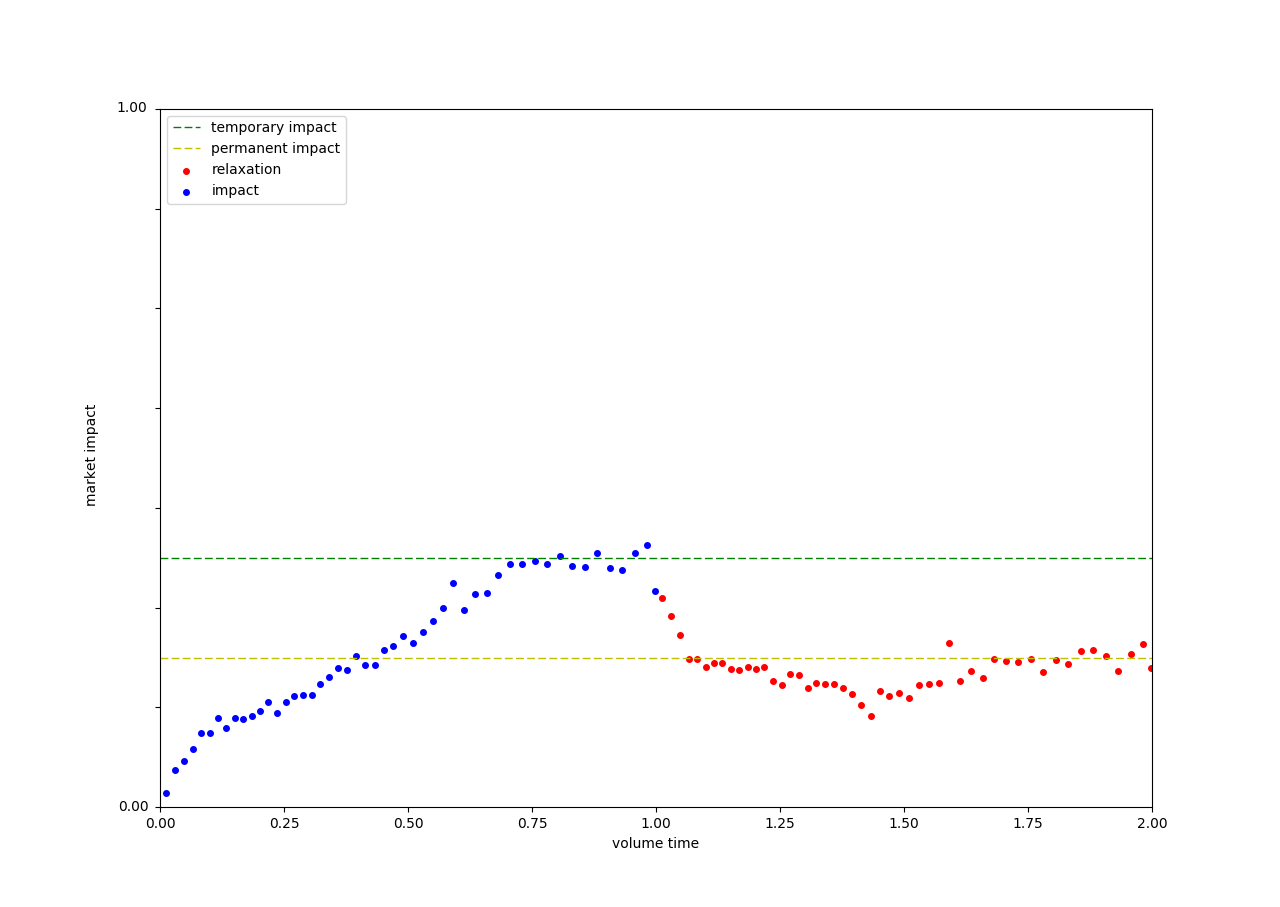}
\captionof{figure}{Market impact dynamics with relaxation in the case of the execution metaorders (set: $\Omega_{30}$, 27 710 metaorders, temporary impact: 0.37, permanent impact: 0.24).}
\label{dynamics relaxation execution 30}
\end{center}
\end{figure}

At first glance, see Figures \ref{dynamics relaxation execution}, \ref{dynamics relaxation execution 10} and \ref{dynamics relaxation execution 30}, the dynamic seems fairly similar to that obtained in the aggressive case, with a slightly rougher curve due to a much smaller data set for the execution metaorders: There is an increasing, concave temporary impact curve and a decreasing, convex relaxation curve that finishes at about two-thirds of the peak impact. However, the monotonicity and convexity of the relaxation curve are less pronounced. Two possible explanations come to mind. The first is that the relaxation is faster because the temporary impacts are less important and therefore, a return to equilibrium is easier. In this hypothesis we can consider the end of the relaxation curve as an other metaorder starting in the market. The second hypothesis is that this is a noisy artifact due to a smaller data set.

\section{Square-Root Law}
\label{square root law}

In this section we are interested in what is commonly called the \textit{Square-Root Law}. The \textit{Square-Root Law} is the fact that the impact curve should not depend on the duration of the metaorder. Indeed, almost all studies now agree on the fact that the impact
is more or less close to be proportional to the square root of the volume executed. However, the so-called \textit{Square-Root Law} states much more than that. It basically claims that the market impact does not depend on the metaorder duration. This last point remains a controversial matter: does the market impact depends on the metaorder duration or not.

\begin{figure}[H]
\begin{center}
\includegraphics[scale=0.46]{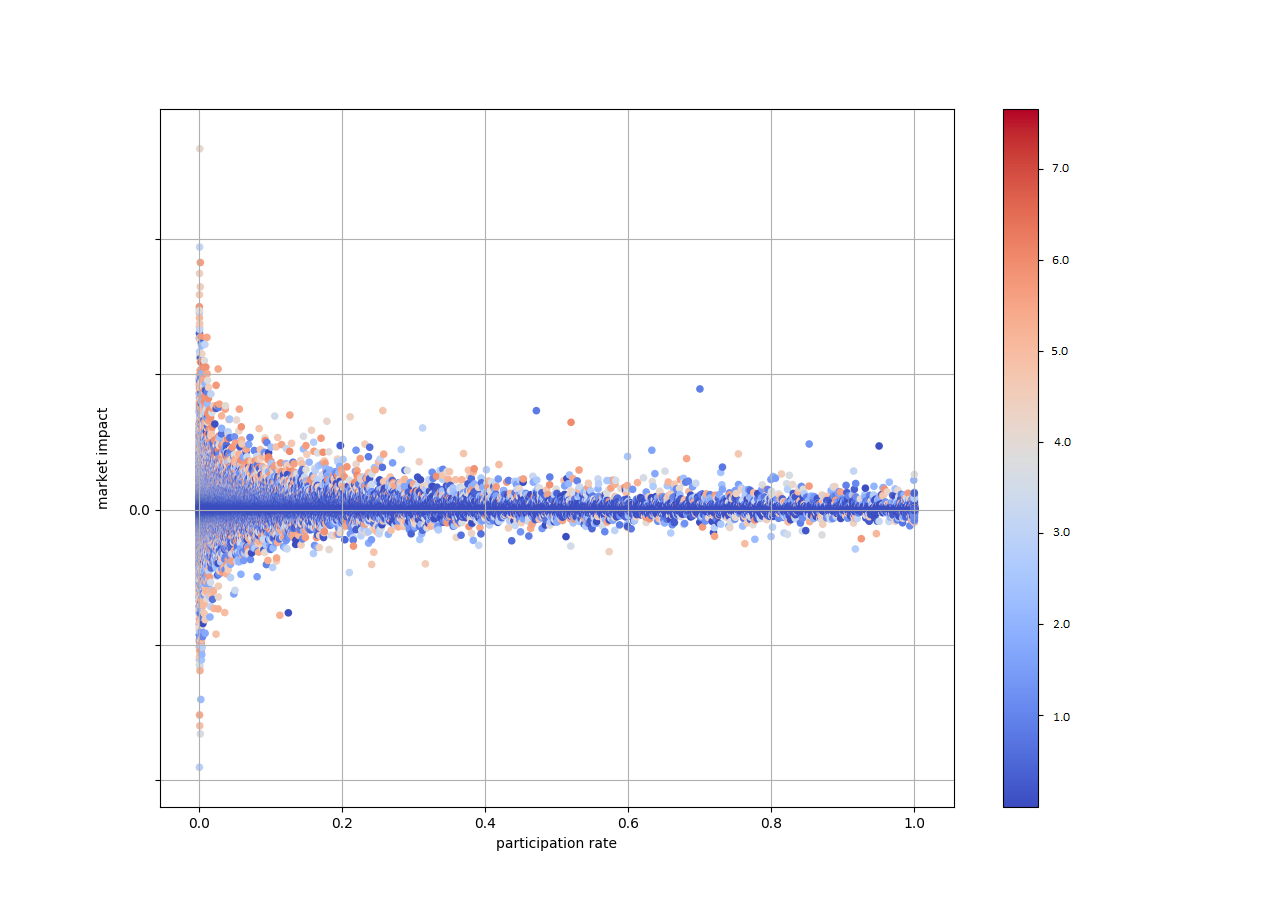}
\captionof{figure}{Impact against participation rate, the duration of the order is increasing from dark blue to dark red in the case of the \textit{aggressive} metaorders.}
\label{square root law aggressive}
\end{center}
\end{figure}

\begin{figure}[H]
\begin{center}
\includegraphics[scale=0.46]{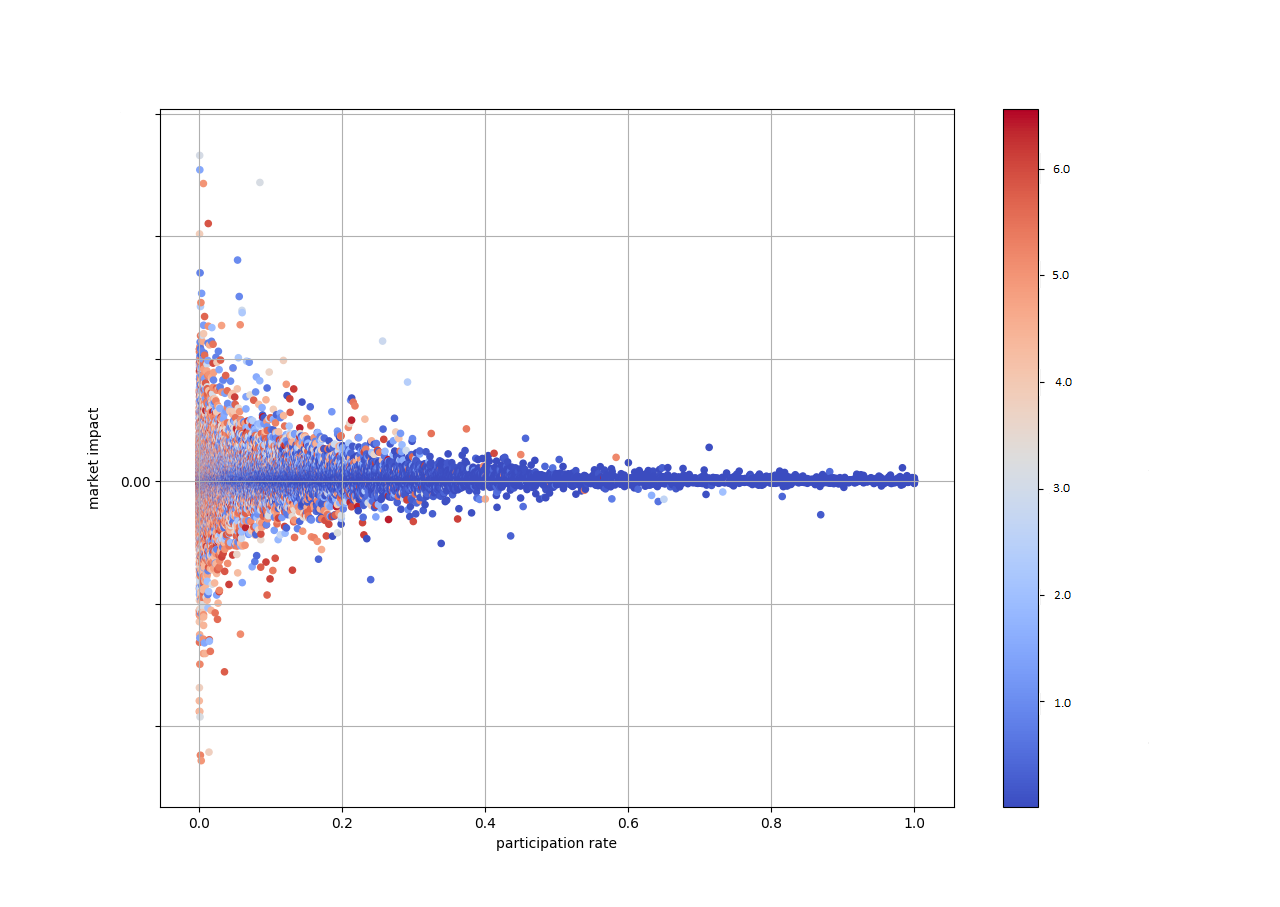}
\captionof{figure}{Impact against participation rate, the duration of the order is increasing from dark blue to dark red in the case of the \textit{execution} metaorders.}
\label{square root law execution}
\end{center}
\end{figure}

At first glance, it does not seem clear that the market impact of a metaorder should depend solely on its size. In the industry, several models are based, at the first order, on a theoretical curve of the form  $\sigma \sqrt{\displaystyle\frac{Q}{V}}$ with $\sigma$ a volatility factor and $\displaystyle\frac{Q}{V}$ a participation rate factor \textbf{over the metaorder time scale}, see e.g. \cite{brokmann2015slow}. Considering such a local participation rate already introduces a duration effect.
Nevertheless, a scatter plot of the impacts in terms of the local participation rates (cf. Figures \ref{square root law aggressive} and \ref{square root law execution})) does not highlight any additional dependency on the duration, apart from that already built in the participation rate. What one can see is that, for the same participation rate, the dispersion of market impact is higher for longer metaorders. This is simply due to the diffusive nature of the prices that creates a noisier impact.

\section{Fair Pricing}
\label{fair pricing}

In this final, short section, the fair pricing condition for metaorders is examined. In the interest of homogeneity, we consider normalized prices, so that for every stock and at every time, the value of the price is $1$ whenever the execution of a metaorder begins.

\begin{figure}[H]
\begin{center}
\includegraphics[scale=0.43]{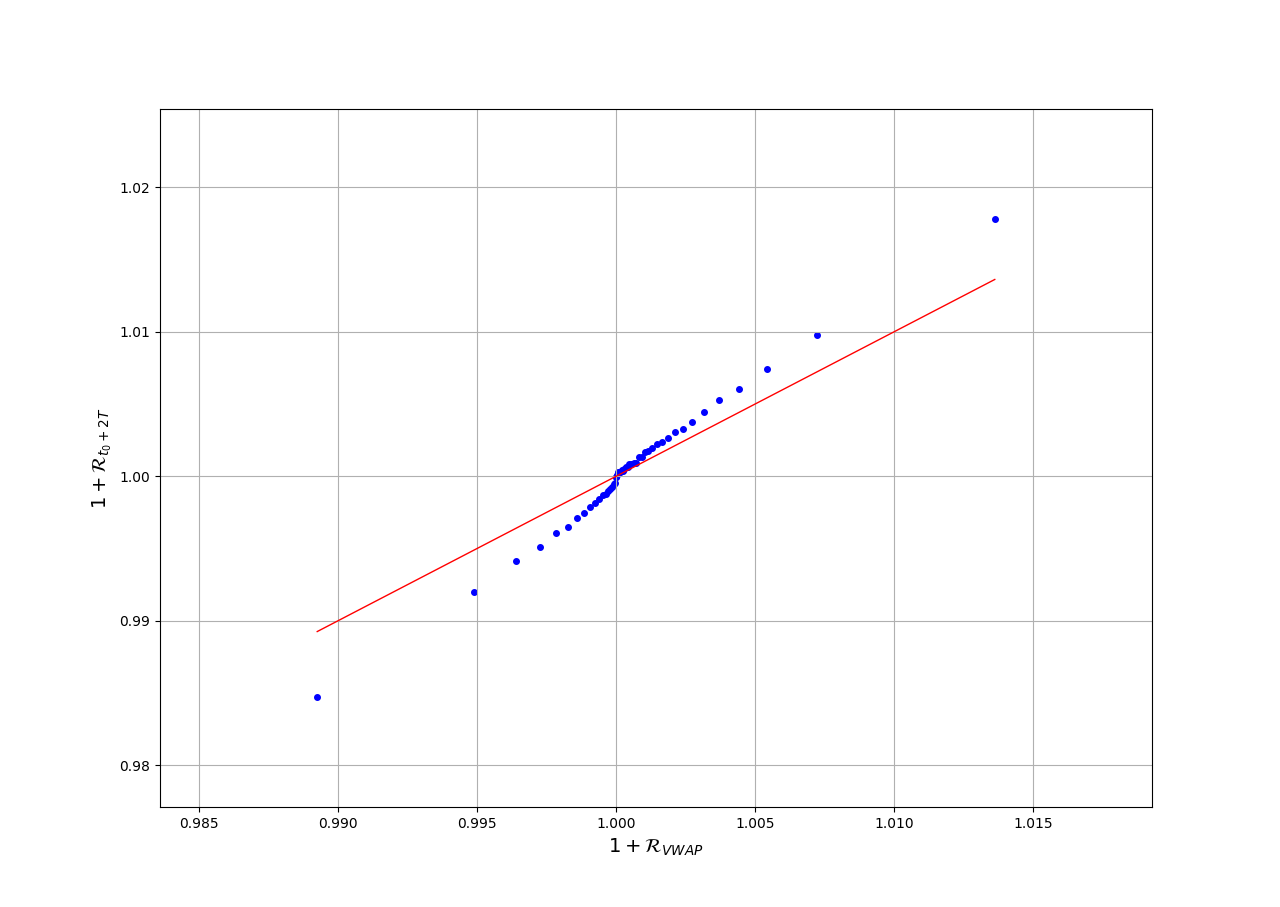}
\captionof{figure}{Fair pricing condition in the case of the aggressive metaorders}
\label{fair pricing aggressive}
\end{center}
\end{figure}

\begin{figure}[H]
\begin{center}
\includegraphics[scale=0.43]{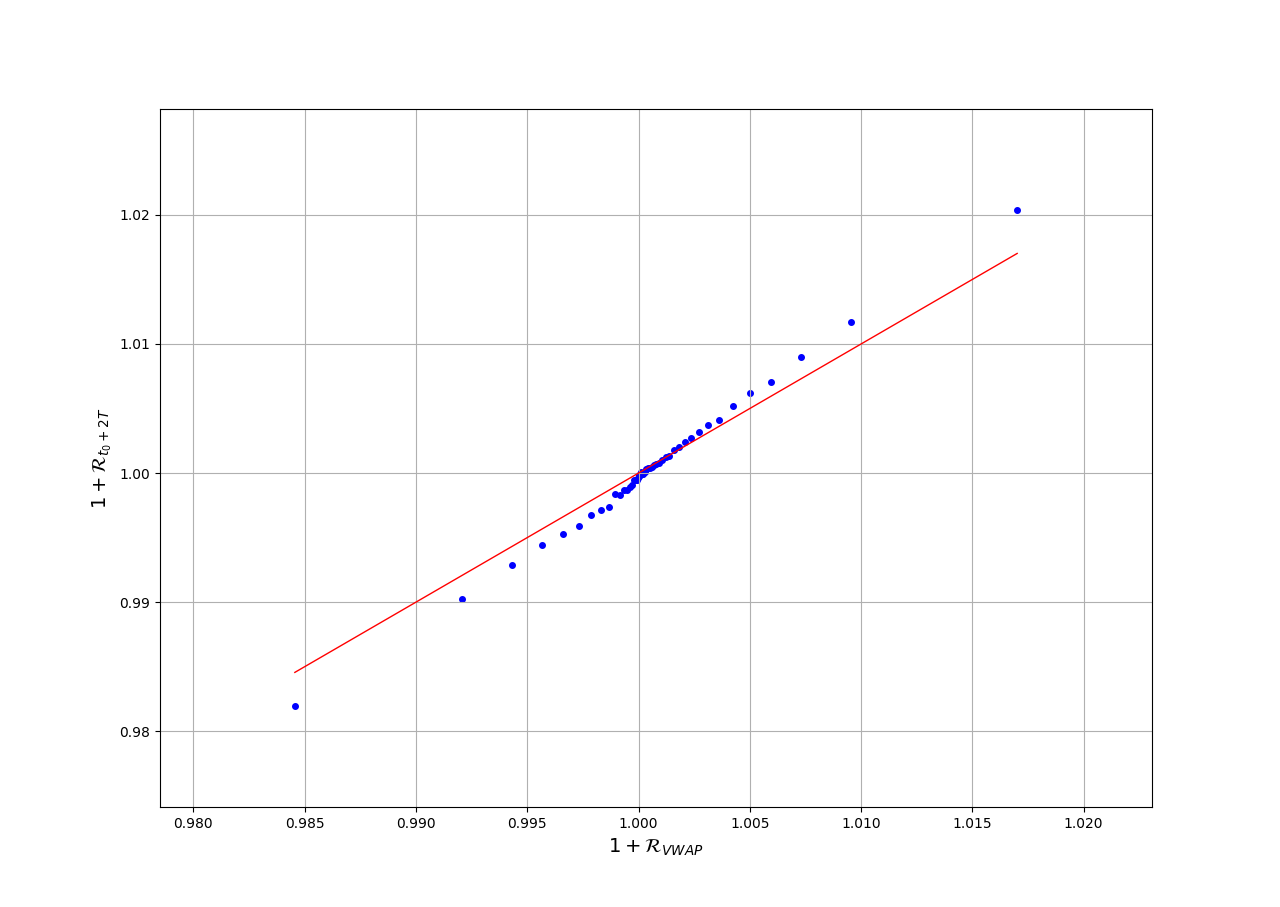}
\captionof{figure}{Fair pricing condition in the case of the execution metaorders}
\label{fair pricing execution}
\end{center}
\end{figure}

First of all, the VWAP of a metaorder $\omega$ is defined by 
$$ \mathcal{P}_{VWAP}(\omega) = \displaystyle\frac{\displaystyle\sum_{i=0}^{N-1}{Q_i(\omega) P_{t_i(\omega)}(\omega)}}{Q(\omega)}, $$
where $t_0(\omega), ..., t_{N-1}(\omega)$ represent the transaction times of the metaorder $\omega$ and $Q(\omega) = \displaystyle\sum_{i=0}^{N-1}{Q_i(\omega)}$. Hence, we want to compare $1 + \mathcal{R}_{VWAP} = \displaystyle\frac{\mathcal{P}_{VWAP}}{P_{t_0}}$ with $1 + \mathcal{R}_{t_0 + 2T} = \displaystyle\frac{P_{t_0 + 2T}}{P_{t_0}}$. The red lines in Figures \ref{fair pricing aggressive} and \ref{fair pricing execution} represent the perfect fair pricing condition, corresponding to the identity $1 + \mathcal{R}_{VWAP} = 1 + \mathcal{R}_{t_0 + 2T}$.

We conclude from Figures \ref{fair pricing aggressive} and \ref{fair pricing execution} that the fair pricing condition can reasonably be assumed to hold. It also appears that, the greater the absolute price variations, the more one moves away from the perfect fair pricing condition. This last point was predictable: high variations are generally associated to longer and larger metaorders that are therefore more affected by the diffusive nature of the prices.
Note that, in the model of \cite{farmer2013efficiency}, the discrete Pareto distribution of $N$ of parameter $\beta + 1 = 2.5$ associated with martingale and fair pricing conditions (see Appendix \ref{Farmer}) leads to a relaxation at $2/3$ of the peak impact, a level that is experimentally verified in our study as well as in the already cited papers \cite{bershova2013non} and \cite{zarinelli2015beyond}.

\section{Conclusion}
\label{conclusion}

The work presented here is an empirical study of a large set of metaorders executed using limit orders in the European equity markets. A new algorithmic definition of a metaorder has been proposed. The statistical results show a good agreement with some predictions of the market impact model in \cite{farmer2013efficiency} in the case of limit orders.

Our study contains two distinct subgroups of orders: a set of aggressive limit orders, and a database of execution strategies predominantly composed of passive limit orders. In both cases, the analysis shows that the temporary market impact is increasing and concave, and it also confirms that the length distribution of metaorders follows a Pareto distribution of parameter $\beta + 1 $ with $\beta \approx 1.4$ in the case of the aggressive metaorders and $\beta \approx 1.8$ in the case of the execution metaorders.
As for the relaxation phase, a convex, decreasing functional form is obtained, as mentioned in \cite{bacry2015market} and other related studies. More precisely, the price reversion after the completion of a trade yields a permanent impact such that its ratio to the maximum impact observed at the last fill is roughly $2/3$ as predicted in the article of \cite{farmer2013efficiency} and highlighted empirically \cite{bershova2013non}.

Finally, we have shown that the square-root law and also the fair pricing condition seem to be empirically verified: the VWAP of a meatorder is equal to the final price of the security after the relaxation phase is over.

\newpage
\bibliographystyle{apalike}
\bibliography{bibliography}

\newpage
\appendix
\section{The Market Impact Model of \cite{farmer2013efficiency}}
\label{Farmer}

The main results of \cite{farmer2013efficiency} are recalled. The central goal of the model is to understand the way order splitting affects market impact.

\subsection{Model description}

\begin{itemize}
    \item A filtered probability space $(\Omega, \mathcal{F}, (\mathcal{F}_t), \mathbb{P})$ is given.
    \item At $t = 0$, before the opening of the market, the $\mathcal{K}$ long-term traders have a common information signal $\alpha$ and each trader $k  = 1,..., \mathcal{K}$ formulate an order and submit it to the algorithmic trading firm that bundles them together into a metaorder which will be executed \textbf{in lots of equal size}.
    \item There are $N$ auctions following each other $t = 1,...,N$, $N$ representing the number of orders necessary to execute fully the metaorder is bounded, $N \leq M$.
    \item At $t = N+1$, corresponding to the relaxation, the last instant in the game, i.e. the instant after the metaorder is fully executed, is announced with the final price $\widetilde{X}_N = G(X_0, \alpha, \eta_1, ..., \eta_N)$, $N > 0$, where $X_0$ is the initial price and $G$ a function whose form is not important. \textbf{We will use the tilde notation to refer to a quantity depending on $(\eta_t)$}, where $(\eta_t)$ is a zero mean i.i.d random process modeling market noise.
\end{itemize}

\subsection{Notations}

\begin{itemize}
    \item As regards statistical averages, $\widetilde{S_t}$ denotes a specific realization of transaction prices, whereas $S_t$ stands for an average price over the signal $\eta_i$. Likewise, $\widetilde{X}_N$ is a specific realization of the final price at the end of the metaorder whereas $X_N$ is an average price over the signal $\eta_i$;
    \item The final price averaging over $\eta_t$  is denoted by $X_N$ i.e $X_N = \mathbb{E}[\widetilde{X}_N]$;
    \item $X_{t-1} = S_{t-1}-R_{t-1}^{-}, t = 2,...,M$ and $X_M = S_M$ thus $R_{M}^{-} = 0$, $M$ corresponding to the end of the market session;
    \item $S_t = X_0 + \displaystyle\sum_{i=0}^{t-1} R_{i}^{+}, t = 1,...,M$, where $R_{t}^{+} = S_{t+1} - S_t$ and $R_{t}^{-} = S_t - X_t$ are the corresponding \text{incremental average impacts};
    \item $\mathcal{I}_t = S_t - X_0$ is the \textbf{average immediate impact} at $t$ and $I_N = X_N - X_0$ denotes the \textbf{average permanent impact};
    \item Let $m \in \{0,1\}$ an indicator variable where $m = 1$ if the metaorder is present and $m = 0$ if it is absent
    \item We will consider $p_t = \mathbb{P}(t \leq N < t+1 \, |\, m = 1)$ at each period $t$ and $\mathcal{P}_t$ the probability that the metaorder will continue given that it is still active at timestep $t$, i.e.
    \begin{center}
    $\mathcal{P}_t = \displaystyle\frac{\displaystyle\sum_{i \geq t+1}{p_i}}{\displaystyle\sum_{i \geq t}{p_i}} = \mathbb{P}(N \geq t+1|N \geq t, m = 1).$
    \end{center}
\end{itemize}

The possible price paths for a buy are summarized in the graph below:
\begin{center}
\includegraphics[scale = 1.0]{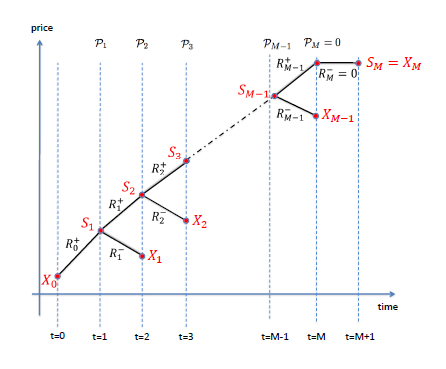}
\captionof{figure}{\textit{Tree of possible price paths for a buy metaorder for different sizes N.}} (extracted from \cite{farmer2013efficiency})
\label{fig1}
\end{center}

In Figure \ref{fig1}, the metaorder is supposed to be present and only expected price paths, averaged over the day trader's information (which is why the notation does not include tildes), are shown. The price is initially $X_0$, after the first lot is executed it is $S_1 = X_0 + R_0^{+}$. If $N = 1$ it is finished and the price reverts to $X_1 = S_1 - R_1^{-}$, but if $N > 1$ another lot is executed and it rises to $S_2 = S_1 + R_1^{+}$. This proceeds similarly until the execution of the metaorder is completed. At any given point the probability that the metaorder has size $N > t$, i.e. that the order continues, is $\mathcal{P}_t$. A typical price path (rather than the expected price paths shown here) subject to a large day trader's noisy information signal would look more like a random walk with a time-varying drift caused by the impact of the metaorder.

\subsection{Martingale condition}
In this section, $(\widetilde{S}_t)$ is supposed to be a martingale: $\mathbb{E}[\widetilde{S}_{t+1}|\mathcal{F}_t] = \widetilde{S}_t$.
Therefore, the following properties hold true:
\begin{itemize}
    \item $\mathbb{E}[\widetilde{S}_{t+1}|\mathcal{F}_t] = \mathbb{P}(m = 1|\mathcal{F}_t) \mathbb{E}[\widetilde{S}_{t+1}|\mathcal{F}_t, m = 1] + \mathbb{P}(m = 0|\mathcal{F}_t) \mathbb{E}[\widetilde{S}_{t+1}|\mathcal{F}_t, m = 0]$, and:
    $$ \mathbb{P}(m=1|\mathcal{F}_t) = \mu_{t}' $$
    considering that $\mu_{t}'$ is the market maker's best estimate of the probability that the metaorder is present;
    \item For the price in the next period, it is necessary to average over three possibilities :
\begin{align*}
    \mathbb{E}[\widetilde{S}_{t+1}|\mathcal{F}_t, m = 1] &= \mathbb{P}(N = t|\mathcal{F}_t, m = 1) \mathbb{E}[\widetilde{S}_{t+1}|\mathcal{F}_t, m = 1, N = t] \\&+ \mathbb{P}(N > t|\mathcal{F}_t, m = 1) \mathbb{E}[\widetilde{S}_{t+1}|\mathcal{F}_t, m = 1, N > t]
\end{align*}
with:
	\begin{itemize}
		\item $\mathbb{E}[\widetilde{S}_{t+1}|\mathcal{F}_t, m = 1, N > t] = \widetilde{S_t} + R_{t}^{+}$
    	\item $\mathbb{E}[\widetilde{S}_{t+1}|\mathcal{F}_t, m = 1, N = t] = \widetilde{S_t} - R_{t}^{-}$
    	\item $\mathbb{E}[\widetilde{S}_{t+1}|\mathcal{F}_t, m = 0] = \widetilde{S}_t$
   		\item $\mathbb{P}(N > t|\mathcal{F}_t, m = 1) = \mathbb{P}(N \geq t+1|N \geq t, m=1) = \mathcal{P}_t$.
	\end{itemize}
\end{itemize}

thus:
    \begin{center}
    $\widetilde{S_t} = \mu_{t}'(\mathcal{P}_t(\widetilde{S_t} + R_{t}^{+}) + (1 - \mathcal{P}_t)(\widetilde{S_t} - R_{t}^{-})) + (1 - \mu_{t}')\widetilde{S_t}$\\
    $\mathcal{P}_t R_{t}^{+} - (1 - \mathcal{P}_t) R_{t}^{-} = 0$
    \end{center}

Since $(\widetilde{S_t})$ is a martingale, there holds:
\begin{prop} ~\\
$$ \textcolor{red}{\displaystyle\frac{R_{t}^{+}}{R_{t}^{-}} = \displaystyle\frac{1 - \mathcal{P}_t}{\mathcal{P}_t}, t \geq 2}. $$
\end{prop}

\subsection{Fair pricing condition}
The martingale condition derived in the previous section only sets the value of the ratio $R_{t}^{+}/R_{t}^{-}$ at each auction $t$. Another condition is required to derive the values of $R_{t}^{+}$ and of $R_{t}^{-}$ and therefore, to obtain the expression for the immediate and the permanent impact.
\begin{itemize}
  \item \textbf{The fair pricing condition} states that for any $N$,
  \begin{center}
  \textcolor{red}{$\pi_N = \displaystyle\frac{1}{N} \displaystyle\sum_{t=1}^{N} {S_t} - X_N = 0.$}
  \end{center}
  This implies that we have
  \begin{center}
  $ \textcolor{red}{I_N = \displaystyle\frac{1}{N} \displaystyle\sum_{t=1}^{N} {\mathcal{I}_t}.} $
  \end{center}
\end{itemize}

Assuming that the martingale condition holds for $t = 1,...,M$ and the fair pricing condition holds for $t = 2,...,M-1$ leads to a system of $2M -2$ homogeneous equations with $2M-1$ unknowns, so we choose $R_1^{+}$ as an undetermined constant.
\begin{prop}
The system of martingale conditions and fair pricing conditions has the solution
$$ \textcolor{red}{R_t^{+} = \displaystyle\frac{1}{t}\displaystyle\frac{1-\mathcal{P}_t}{\mathcal{P}_t}\displaystyle\frac{1}{\mathcal{P}_1\mathcal{P}_2...\mathcal{P}_{t-1}}R_1^{+}, t \geq 2} $$
and
$$ \textcolor{red}{R_t^{-} = \displaystyle\frac{1}{t}\displaystyle\frac{1}{\mathcal{P}_1\mathcal{P}_2...\mathcal{P}_{t-1}}R_1^{+}, t \geq 2.} $$
\end{prop}

\begin{coro}
The immediate impact is
$$ \textcolor{red}{\mathcal{I}_t = S_t - X_0 = R_{0}^{+} + R_{1}^{+}\left(1 + \displaystyle\sum_{k=2}^{t-1} \displaystyle\frac{1}{k} \displaystyle\frac{1-\mathcal{P}_k}{\mathcal{P}_k} \displaystyle\frac{1}{\mathcal{P}_1...\mathcal{P}_{k-1}}\right), t \geq 2.} $$
\end{coro}

Since $X_N - X_0 = X_N - S_N + S_N - X_0 = \mathcal{I}_N - R_N^{-}$, we have

\begin{coro}
The permanent impact is
$$ \textcolor{red}{I_N = R_{0}^{+} + R_{1}^{+}\left(1 + \displaystyle\sum_{k=2}^{N-1} \displaystyle\frac{1}{k} \displaystyle\frac{1-\mathcal{P}_k}{\mathcal{P}_k} \displaystyle\frac{1}{\mathcal{P}_1...\mathcal{P}_{k-1}} -\displaystyle\frac{1}{N}\displaystyle\frac{1}{\mathcal{P}_1...\mathcal{P}_{N-1}}\right), N \geq 2.} $$
\end{coro}

\subsection{Dependence on the metaorder size distribution} \label{Dependence on the metaorder size distribution}
According to \cite{farmer2013efficiency} which cite a host of other relevant studies, there is a considerable evidence that - in the large size limit and for most major equity markets - the metaorder size $V$ is distributed as $\mathbb{P}(V > v) \sim v^{-\beta}$, with $\beta \approx 1.5$.
\begin{itemize}
  \item $\mathbb{P}(V > v-1) - \mathbb{P}(V > v) \sim v^{-\beta}\left(\left(1-\displaystyle\frac{1}{v}\right)^{-\beta} - 1\right) \approx \displaystyle\frac{\beta}{v^{\beta + 1}}$: an exact zeta distribution for all $n \geq 1$ is considered,
  \begin{center}
  $p_n = \mathbb{P}(N=n|m=1) = \displaystyle\frac{1}{\zeta(\beta +1)}\displaystyle\frac{1}{n^{\beta +1}} $
  \end{center}
  so that
  \begin{center}
  $\mathcal{P}_t = \displaystyle\frac{\zeta(1+\beta,t+1)}{\zeta(1+\beta,t)} \approx \displaystyle\frac{\displaystyle\int_{t+1}^{+\infty} \displaystyle\frac{\mathrm{d}x}{x^{\beta +1}}}{\displaystyle\int_{t}^{+\infty} \displaystyle\frac{\mathrm{d}x}{x^{\beta +1}}} = \left(1+\displaystyle\frac{1}{t}\right)^{-\beta} \sim 1 -\displaystyle\frac{\beta}{t}$;
  \end{center}
  \item As a consequence,
  \begin{center}
  $R_t^{+} = \displaystyle\frac{1}{t^{2+\beta}}\displaystyle\frac{\zeta(1+\beta)}{\zeta(1+\beta,t)\zeta(1+\beta,t+1)}R_1^{+} \sim \displaystyle\frac{1}{t^{2-\beta}}$;
  \end{center}
  \vspace{0.5cm}
  \item Thus, the immediate impact $\mathcal{I}_t$ behaves asymptotically for large $t$ as 
  \begin{center}
  $\mathcal{I}_t \sim \bigg\{
    \begin{array}{rl}
    t^{\beta-1} &for \,\beta \ne 1 \\
        log(t+1) &for \,\beta = 1
    \end{array}
    ;$
 \end{center}
  \item Recalling the fact that the fair pricing gives $I_N = \displaystyle\frac{1}{N}\displaystyle\sum_{t=1}^{N} \mathcal{I}_t$, there holds
  \begin{center}
  $I_N \sim \displaystyle\frac{1}{N}\displaystyle\int_{0}^{N} x^{\beta -1}\,\mathrm{d}x = \displaystyle\frac{1}{\beta}N^{\beta -1}$.
  \end{center}
\end{itemize}

Finally

\begin{prop}~\\
$$\textcolor{red}{\displaystyle\frac{I_N}{\mathcal{I}_N} = \displaystyle\frac{1}{\beta}.}$$
\end{prop}

For further use, one can observe that for a value of $\beta = 1.5$, we have $\displaystyle\frac{I_N}{\mathcal{I}_N} = \displaystyle\frac{2}{3}$.

\end{document}